\documentclass[aps,prx,amsmath,amssymb,twocolumn,superscriptaddress,notitlepage]{revtex4-2}
\usepackage{graphicx}
\usepackage{color}
\usepackage{tikz}
\usepackage{braket}
\usepackage{xcolor}
\usepackage{framed}
\definecolor{burntorange}{rgb}{0.8, 0.33, 0.0}\usepackage[colorlinks,citecolor=blue,linkcolor=red,urlcolor=blue]{hyperref}
\hypersetup{
  pdfkeywords={},
  pdftitle={Efficient Information Retrieval for Sensing via Continuous Measurement},
  }

\makeatletter

\definecolor{DYblue}{RGB}{1,100,0}

\colorlet{shadecolor}{blue!10}

\begin{document}

\title{Efficient Information Retrieval for Sensing via Continuous Measurement}

\author{Dayou Yang}
\author{Susana F. Huelga}
\author{Martin B. Plenio}

\affiliation{
Institut f\"ur Theoretische Physik and IQST, Universit\"at Ulm,
Albert-Einstein-Allee 11, D-89069 Ulm, Germany}

\date{\today}

\begin{abstract}
Continuous monitoring of driven-dissipative quantum optical systems is a crucial element in the implementation of quantum metrology, providing essential strategies for achieving highly precise measurements beyond the classical limit. In this context, the relevant figure of merit is the quantum Fisher information of the radiation field emitted by the driven-dissipative sensor. Saturation of the corresponding precision limit as defined by the quantum Cram\'er-Rao bound is typically not achieved by conventional, temporally local continuous measurement schemes such as counting or homodyning. To address the outstanding open challenge of efficient retrieval of the quantum Fisher information of the emission field, we design a novel continuous measurement strategy featuring temporally quasi-local measurement bases as captured by matrix product states. Such measurement can be implemented effectively by injecting the emission field of the sensor into an auxiliary open system, a `quantum decoder' module, which `decodes' specific input matrix product states into simple product states as its output field, and performing conventional continuous measurement at the output. We devise a universal recipe for the construction of the decoder by exploiting the time reversal transformation of quantum optical input-output channels, thereby establishing a universal method to achieve the quantum Cram\'er-Rao precision limit for generic sensor designs based on continuous measurement. As a by-product, we establish an effective formula for the evaluation of the quantum Fisher information of the emission field of generic driven-dissipative open sensors. We illustrate the power of our scheme with paramagnetic open sensor designs including linear force sensors, fibre-interfaced nonlinear emitters, and driven-dissipative many-body sensors, and demonstrate that it can be robustly implemented under realistic experimental imperfections. 
\end{abstract}
\maketitle

\section{Introduction}
\label{sec:intro}
\begin{figure*}[t!]
\centering{} \includegraphics[width=1\textwidth]{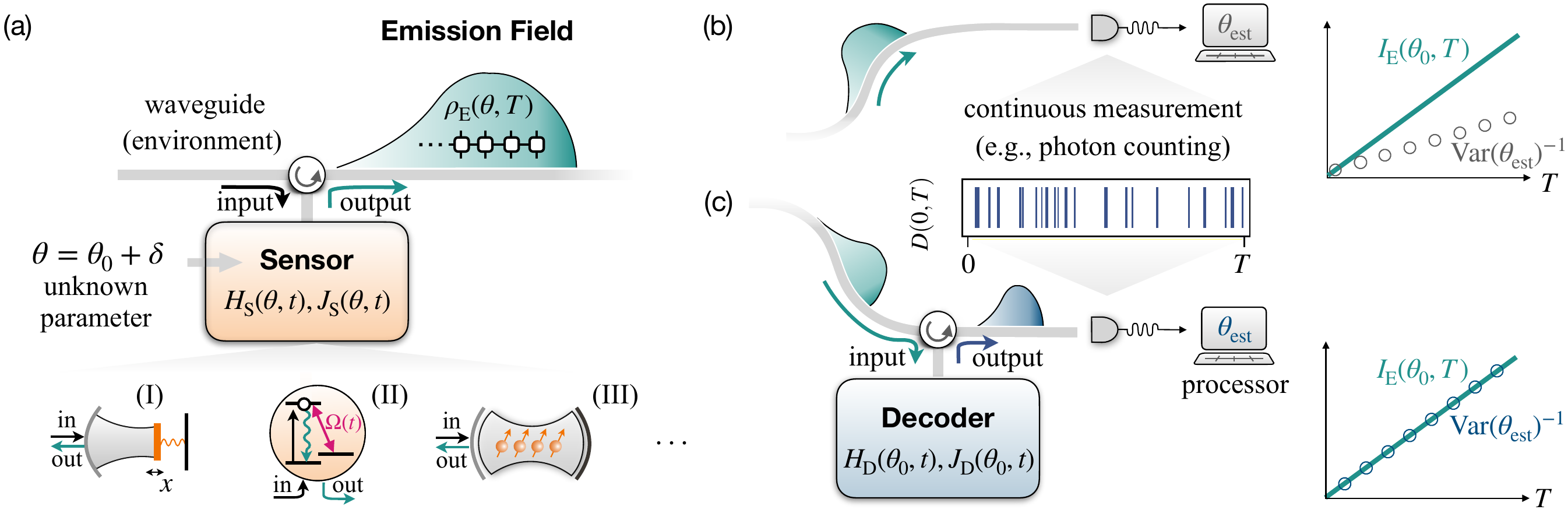} 
\caption{Schematic of continuous-measurement based sensing technology and our efficient information retrieval scheme. (a) Continuous-measurement based sensing consists of driving an open dissipative sensor with an input field (e.g., laser) and detecting the emission field of the sensor continuously in time. A generic driven-dissipative sensor can be parameterized by a parameter-encoded Hamiltonian $H_{\rm S}(\theta,t)$ and jump operator $J_{\rm S}(\theta,t)$. Its emission field in the time window $[0,T]$ is described by a correlated multi-photon state $\rho_{\rm E}(\theta, T)$ with a matrix-product structure. Paradigmatic examples of this class of sensors include, e.g., (I) linear force sensors (II) nonlinear quantum emitters and (III) driven-dissipative quantum many-body sensors. (b) Conventional sensing schemes detect the emission field via time-local measurements, e.g., counting and homodyning. Nevertheless, the complex matrix-product structure of the emission field typically renders these measurements ineffective to retrieve the full quantum Fisher information of the emission field, $I_{\rm E}(\theta_0,T)$, with $\theta_0$ the prior knowledge of $\theta$. This severely limits the sensing precision, as quantified by the inverse variance of the estimation ${\rm Var}(\theta_{\rm est})^{-1}$. (c) We develop an efficient information retrieval strategy by injecting the emission field of the sensor into a quantum decoder module, whose driven-dissipative evolution maps the sensor emission field $\rho_{\rm E}(\theta_0, T)$ to a product state (e.g., vacuum). Subsequent counting of the output field of the decoder, as we will show, accomplishes measurement of the sensor emission field in an optimal matrix-product basis, capable of achieving the optimal precision of the sensor design. We provide a universal recipe for the construction of the decoder parameters, Eq.~\eqref{eq:H_J_general_solution}, applicable to generic, potentially nonlinear and time-dependent sensor design, and demonstrate it using the examples (I-III).}
\label{fig:fig1} 
\end{figure*}

An ongoing pursuit in the field of quantum sensing is the development of methods that achieve the fundamental sensitivity limit allowed by quantum mechanics. Given a sensor design, such a limit is set by the quantum {Cr\'amer-Rao} bound (QCRB)~\cite{PhysRevLett.72.3439} for the mean-square error in parameter estimation: ${\rm Var}[\theta_{\rm est}]\geq 1/{K I(\theta)}$, where $K$ is the number of sensing repetitions, and $I(\theta)$ is the quantum Fisher information (QFI). Achieving the QCRB requires full retrieval of the QFI, which is possible only via an optimal quantum measurement strategy---a strategy that minimizes the influence of the intrinsic quantum noise underlying the sensor design.

A large class of important sensor technology is based on \emph{continuous measurement}~\cite{Braginsky1992,carmichael1993open,RevModPhys.70.101,gardiner2004quantum,Wiseman,gardiner2015quantum} which, as illustrated in Fig.~\ref{fig:fig1}(a), consists of driving an open dissipative sensor with lasers (microwaves) and detecting the emission field of the sensor continuously, e.g., via time-resolved photon counting or homodyning.  Prominent examples include gravitational wave detectors~\cite{PhysRevLett.116.061102,Danilishin:2012wp} and optomechanical force sensors~\cite{RevModPhys.82.1155,RevModPhys.86.1391,Barzanjeh:2022ve}, whose key element is a driven-dissipative mechanical oscillator (e.g., a cavity mirror); and atomic gas magnetometers~\cite{Moller:2017ta,Colangelo:2017wi}, which consist of the collective spins of a laser-driven atomic ensemble---these are representatives of \emph{linear} driven-dissipative sensors. In contrast, rapid progress towards the realization of \emph{nonlinear}, potentially \emph{many-body} and \emph{interacting} driven-dissipative sensors substantially broadens the class of sensor technology based on continuous measurement. These advances are highlighted by the recent experimental achievements in the efficient interface between tailored photonic structures, e.g. low-loss optical fibres (waveguides)~\cite{doi:10.1063/1.4838696,PhysRevLett.110.243602,Hacker:2016ti,Landig:2018ta,Mirhosseini:2019uf} and nanophotonic platforms~\cite{doi:10.1073/pnas.1603788113,doi:10.1126/science.aaj2118,Lodahl:2017vb,doi:10.1126/science.abi9917}, with driven-dissipative quantum optical systems ranging from single nonlinear emitters~\cite{Northup:2014tt,RevModPhys.87.347,RevModPhys.87.1379}, via molecules~\cite{Shalabney:2015ve}, to intracavity Bose-Einstein condensates~\cite{RevModPhys.85.553,Dreon:2022wd,doi:10.1126/science.abo3382} with increasing complexity.

Along with the experimental advances, manifold theoretical investigations are devoted to understanding the performance of generic sensor designs based on continuous measurement. Progress include, e.g., the evaluation of the QCRB~\cite{PhysRevLett.106.090401,PhysRevLett.112.170401,2015JPhA...48J5301C}, signal processing~\cite{Mabuchi_1996,PhysRevA.64.042105,PhysRevA.89.052110,PhysRevA.94.032103}, the impact of noise~\cite{Albarelli_2017,Albarelli2018restoringheisenberg,PhysRevLett.125.200505} and criticality enhanced sensitivity~\cite{PhysRevA.93.022103,PRXQuantum.3.010354}. Despite the progress, a severe obstacle to the improvement of the sensor performance has been revealed~\cite{Mabuchi_1996,PhysRevLett.106.090401,PhysRevLett.112.170401,2015JPhA...48J5301C,PhysRevA.64.042105,PhysRevA.89.052110,PhysRevA.94.032103,Albarelli_2017,Albarelli2018restoringheisenberg,PhysRevLett.125.200505,PhysRevA.93.022103,PRXQuantum.3.010354}---conventional \emph{time-local} continuous measurement, e.g., counting or homodyning, typically retrieves only a small portion of the QFI of the emission field, thus fail to achieve the QCRB of \emph{almost all} sensor designs, as illustrated in Fig.~\ref{fig:fig1}(b). This  originates from the fact that the parameter to be sensed is hidden in the emission field of the sensor which, as a result of the sensor evolution, can be \emph{nonclassical}, \emph{complex} and \emph{strongly correlated in time}, forbidding its full retrieval via conventional time-local measurement schemes.

Consequently, an outstanding challenge naturally emerges---what is the optimal measurement strategy for achieving the QCRB of a generic driven-dissipative quantum sensor, and how to implement such measurement effectively? Despite its fundamental importance to the development of sensing technology based on continuous measurement, this question is largely open---except for the special case of linear sensors~\cite{PhysRevLett.116.061102,Danilishin:2012wp,RevModPhys.82.1155,RevModPhys.86.1391,Barzanjeh:2022ve,Moller:2017ta,Colangelo:2017wi,*[{For recent development for improving the sensitivity of linear sensors, see, e.g., }][{}] PhysRevA.93.042121,*PhysRevA.107.012611}, where the linear Heisenberg-Langevin equations of motion reveals backaction evasion techniques~\cite{PhysRevD.65.042001,PhysRevLett.99.110801,PhysRevLett.104.133602,Hertzberg2010}, e.g., the `negative mass' oscillator scheme~\cite{Moller:2017ta,PhysRevLett.102.020501,PhysRevLett.105.123601,PhysRevLett.121.103602,PhysRevLett.121.031101}, for approaching the QCRB, see our discussion below. For generic \emph{nonlinear} and \emph{complex} open sensor designs, however, saturation of the QCRB remains an outstanding open challenge. 

In this manuscript, we overcome this challenge by proposing and developing a novel continuous measurement scheme 
that establishes a \emph{universal} strategy for achieving the QCRB for generic continuous-measurement based sensors---which can be nonlinear and even non-stationary (e.g., subjected to time-dependent driving). Our solution is inspired by the intimate relation between the emission field of generic driven-dissipative emitters and the concept of matrix product states (MPSs)~\cite{RevModPhys.93.045003,PhysRevLett.95.110503,PhysRevLett.105.260401,PhysRevLett.116.093601}---in stark contrast to conventional time-local continuous measurement, our scheme features projective measurement in a temporally \emph{quasi-local}, finitely correlated basis as captured by MPS structures. Such a measurement scheme can be implemented naturally by harnessing the features of quantum optical driven-dissipative dynamics---dependent on the dynamics of the open system, the environment can be mapped from a product (e.g., vacuum) input state to a specific MPS output, and vice versa. Given a generic driven-dissipative quantum sensor, as described by the parameter-encoded (potentially time-dependent) Hamiltonian $H_{\rm S}(\theta,t)$ and jump operator $J_{\rm S}(\theta,t)$, cf. Fig.~\ref{fig:fig1}(a), we can couple its emission field to an auxiliary driven-dissipative system, as illustrated in Fig.~\ref{fig:fig1}(c), {which we call a \emph{quantum decoder} module by drawing analogies to unitary decoding in optimal Ramsey interferometry~\cite{PhysRevX.11.041045,Marciniak:2022aa}}. Dependent on its driven-dissipative evolution, as governed by its Hamiltonian $H_{\rm D}$ and jump operator $J_{\rm D}$, the decoder `decodes' specific input MPSs into simple product (e.g., vacuum) state as its output. Such possibilities was first explored in Ref.~\cite{Stannigel_2012} to construct coherent quantum absorbers of time-stationary emitters for dissipative phase engineering~\cite{PhysRevLett.113.237203,PhysRevA.91.042116}. Here, we expand upon them to encompass also time-dependent scenarios as pertinent to sensing, by exploiting general \emph{time reversal transformation} of quantum optical input-output channels. Remarkably, as we establish below, such a transformation is described mathematically by changing the canonical form of MPSs~\cite{RevModPhys.93.045003}.
As a result of decoding, photon counting at the decoder output accomplishes projective measurement of its input, i.e., the emission field of the sensor, in temporally finite-range correlated MPS basis. We emphasize that such correlation originates from the driven-dissipative quantum evolution of the decoder, in stark contrast to the correlation built upon time-local measurement with classical feedforward (feedback). 

Saturation of the QCRB is achieved by designing the decoder evolution in such a way that the MPSs prior to the decoding corresponds to the optimal measurement basis of the emission field, thus fully retrieving its QFI~\footnote{\label{note1}In the context of sensing via continuous measurement, the relevant QFI typically grows with the measurement time. Correspondingly, Saturation of the QCRB is typically defined in terms of long-time optimum, i.e., the retrieved Fisher information differs from the QFI by an asymptotic constant for long measurement time. We adopt this definition in the present work.}. For a generic (nonlinear) open sensor, the QCRB can only be achieved in the neighborhood of the prior knowledge $\theta_0$ of the unknown parameter---the optimal measurement basis, and correspondingly the required decoder evolution, is dependent on $\theta_0$, $H_{\rm D}=H_{\rm D}(\theta_0,t)$ and $J_{\rm D}=J_{\rm D}(\theta_0,t)$, cf. Fig.~\ref{fig:fig1}(c). We develop a simple yet \emph{universal} recipe for the construction of $H_{\rm D}(\theta_0,t)$ and $J_{\rm D}(\theta_0,t)$, cf., our central formula Eq.~\eqref{eq:H_J_general_solution}, based on the aforementioned time reversal transformation. Remarkably, such a recipe only requires the knowledge of the sensor parameters $H_{\rm S}(\theta_0,t)$ and $J_{\rm S}(\theta_0,t)$, and does not rely on the extensive knowledge of the emission field of the sensor as provided, e.g., via tomography of the field. With the decoder properly tuned, counting its output field and data processing provide us with a QCRB-limited estimation precision in the neighborhood of $\theta_0$, as illustrated in Fig.~\ref{fig:fig1}(c). 

Our information retrieval scheme can be viewed as a universal quantum backaction evasion strategy for generic (linear and nonlinear) driven-dissipative sensors. We illustrate its power using three paradigmatic continuous-measurement based sensor designs at increasing complexity. (I) Linear open quantum sensors~\cite{PhysRevLett.116.061102,Danilishin:2012wp,RevModPhys.82.1155,RevModPhys.86.1391,Barzanjeh:2022ve,Moller:2017ta,Colangelo:2017wi}, for which we show that our decoder module naturally reduces to a displaced `negative mass' oscillator. (II) Nonlinear emitters driven by stationary or time-varying fields, which are emergent platforms for the generation of large-scale multi-photon entangled states~\cite{PhysRevLett.95.110503,PhysRevLett.116.093601,PhysRevLett.120.130501,PhysRevLett.128.010607,Besse:2020vq} for sensing. Remarkably, we demonstrate that by harnessing the highly correlated emission field generated by time-dependent driving fields, our retrieval scheme allows for achieving a Heisenberg-limited QCRB. (III) Driven-dissipative lattice spin models as realizable, e.g., in trapped-ion~\cite{doi:10.1126/science.abg8102,doi:10.1126/science.abk2400} or neutral atom platforms~\cite{Simon:2011vs,Scholl:2021ug,Ebadi:2021wa}, as a representative of the significant experimental progress towards the integration of synthetic many-body systems as sensors~\cite{Goban:2018aa,Marciniak:2022aa,Ding:2022ve}. Besides these paradigmatic examples, we demonstrate that the temporally quasi-local, finitely-correlated nature of our measurement scheme offers it remarkable resilience against experimental imperfections including, e.g., light transmission loss and inaccurate control of the decoder, such that it can be robustly implemented in experiments. As such, our scheme provides a general and practical strategy for improving open quantum sensors towards their ultimate precision limit, with broad applications to diverse platforms across experimental quantum optics. 

Along with the efficient information retrieval scheme, we establish an effective method to evaluate the QFI of the emission field of generic driven-dissipative sensors thanks to our MPS formulation. This provide us with a refined, tighter QCRB for sensors based on continuous measurement, as compared to existing bound~\cite{PhysRevLett.106.090401,PhysRevLett.112.170401,2015JPhA...48J5301C} that are expressed in terms of the global QFI of the sensor and the emission field. The significance of our tighter QCRB is particularly highlighted by nonergodic sensors, i.e., sensors that possess multiple stationary states, for which we show that the two QCRBs can exhibit disparate scaling behavior with respect to the interrogation time. In contrast, for ergodic sensors, i.e., sensors that possess a unique stationary state, we show that despite a finite difference, both QCRBs follow the same linear scaling with respect to the interrogation time. We demonstrate the connection and difference between the two QCRBs using the aforementioned sensor models.

The rest of our manuscript is organized as follows. We introduce a MPS representation of the emission field of generic driven-dissipative sensors in Sec.~\ref{sec:minimalModel}, as the basis to formulate our sensing framework and our retrieval scheme. We then present in Sec.~\ref{sec:eQFI} the framework of sensing via continuous-measurement based on the MPS formulation, highlighting an effective method for the evaluation of the QFI of the emission field of generic driven-dissipative sensors. We establish our new continuous measurement scheme for efficient information retrieval in Sec.~\ref{eq:eff_retrieval}, and demonstrate its power using paradigmatic examples of open quantum sensors in Sec.~\ref{sec:examples}. We discuss the experimental feasibility of our scheme in Sec.~\ref{sec:imperfect}, and conclude in Sec.~\ref{sec:concusion} with a summary of our results and an outlook.

\section{Matrix product state representation of the emission field}
\label{sec:minimalModel}
Central to our sensing framework is the detection of the emission field of open quantum optical systems for sensing. To lay its basis and fix our notation, let us start our discussion with a self-contained introduction to an effective description of the quantum state of the emission field in terms of MPSs, based on familiar quantum optics theory~\cite{carmichael1993open,RevModPhys.70.101,breuer2002theory,gardiner2004quantum,Wiseman,gardiner2015quantum,Rivas_2012}. 

\subsection{Quantum Optical Model}
\label{sec:QOmodel}
We consider the sensor as a quantum optical open system coupled with a one-dimensional bosonic environment (e.g. an optical fiber or a waveguide) representing the input and output channel of the system, cf. Fig.~\ref{fig:fig1}. The joint evolution of the system and the environment in the Schr\"odinger picture is governed by the Hamiltonian (we set $\hbar=1$ hereafter)
\begin{equation}
\label{eq:Hse}
{\cal H}_{\rm SE}={\cal H}_{\rm S}+{\cal H}_{\rm E}+{\cal V}_{\rm SE},
\end{equation}
where ${\cal H}_{\rm S(E)}$ is the Hamiltonian of the system (environment) and ${\cal V}_{\rm SE}$ describes the coupling between them. Here, we have adopted the convention that ${\cal H}$$({H})$ represents a Hamiltonian in the Schr\"odinger (interaction) picture, the latter to be introduced below. Denoting the photonic destruction (creation) operators of frequency $\omega$ as $b(\omega)$ $[b^\dag(\omega)]$, we can express the environment Hamiltonian as
\begin{equation}
\label{eq:HE}
{\cal H}_{\rm E}=\int_{\cal} d\omega \,\omega\, b^\dag(\omega)b(\omega),
\end{equation}
and the system-environment coupling Hamiltonian as
\begin{equation}
\label{eq:VSE}
{\cal V}_{\rm SE}=i\frac{1}{\sqrt{2\pi}}\int_{\bar{\omega}-\cal B}^{\bar{\omega}+\cal B} \! d\omega \,[J_{\rm S}b^\dag(\omega) - J_{\rm S}^\dag b(\omega)].
\end{equation}
Here, $J_{\rm S}$ is the system transition (jump) operator, and we have made the typical assumption in quantum optics~\cite{carmichael1993open,RevModPhys.70.101,breuer2002theory,gardiner2004quantum,Wiseman,gardiner2015quantum,Rivas_2012}, i.e., the bandwidth $2\cal B$ of the weak coupling ${\cal V}_{\rm SE}$ is much smaller than the system transition frequencies centered around $\bar{\omega}$ (the latter lies in the optical frequency range). This allows us to assume that the coupling strength in ${\cal V}_{\rm SE}$ is $\omega$-independent, which is absorbed into $J_{\rm S}$ for notational simplicity, and to neglect counter-rotating terms in ${\cal V}_{\rm SE}$.

To arrive at an effective description of the dynamics, let us transform away the optical frequencies in ${\cal H}_{\rm S(E)}$ by moving to an interaction picture with respect to $\bar{\omega} J^\dag_{\rm S} J_{\rm S} + {\cal H}_{\rm E}$. This results in time-dependent operators for the system and the environment. In particular, this allows us to define the (time-dependent) environmental operator~\cite{carmichael1993open,RevModPhys.70.101,gardiner2004quantum,Wiseman,gardiner2015quantum}
\begin{equation}
b(t)=\frac{1}{\sqrt{2\pi}}\int_{\bar{\omega}-\cal B}^{\bar{\omega}+\cal B} \! d\omega \,b(\omega)e^{-i(\omega-\bar{\omega})t}.
\end{equation}
Correspondingly, we can express the joint Hamiltonian of the system and the environment as
\begin{equation}
\label{eq:HSEint}
H_{\rm SE} = {H}_{\rm S}(t)+i\left[J_S(t) b^\dag(t)-J_S^\dag(t) b(t)\right],
\end{equation}
where ${H}_{\rm S}(t)$ and $J_S(t)$ is the transformed Hamiltonian and jump operator of the system that capture its slow dynamics in the interaction picture. They can be time-dependent (with a characteristic frequency much smaller than ${\cal B}$) for general quantum optical settings, as relevant to the various sensor designs discussed below.

\subsection{Coarse-Grained Description}
\label{sec:coarse-graining}
Following familiar quantum optics theory~\cite{carmichael1993open,RevModPhys.70.101,breuer2002theory,gardiner2004quantum,Wiseman,gardiner2015quantum,Rivas_2012}, we can describe the slow dynamics in the interaction picture by considering a coarse-grained time scale much longer than the inverse bandwidth ${\cal B}^{-1}$ while being much shorter than the time scale of the slow dynamics. This promotes the environmental modes $b(t)$ to quantum noise operators with the white noise bosonic commutation relations $[b(t), b^\dag(t^\prime)]=\delta(t-t')$, allowing us to integrate conveniently the time evolution governed by Eq.~\eqref{eq:HSEint}. 

We hereby define the elementary time increment $\Delta t\gg{\cal B}^{-1}$, and divide the time window $[0,T)$ in terms of $N$ elementary time bins $[n]:=[t_{n-1},t_n)$, $n\in[1,N]$, with $t_n=n\Delta t$ and $T=N \Delta t$. We define for time bin $[n]$ the quantum noise increment
\begin{equation}
\Delta B_{[n]}= \int_{t_{n-1}}^{t_n}\!d\tau \,b(\tau),
\end{equation}
which satisfies the bosonic commutation relation $[\Delta B_{[n]},\Delta B_{[n']}^\dag]=\delta_{nn'}\Delta t$ and therefore defines an individual time-bin mode, for which the Fock basis can be constructed as $|\sigma\rangle_n=\Delta B_{[n]}^{\dag\sigma} |0\rangle_n/\sqrt{\sigma!\Delta t^\sigma}$. We can express the time evolution in the time bin $[n]$ by the expansion of $\exp(-i\int_{t_{n-1}}^{t_{n}}d\tau H_{\rm SE})$ to ${\cal O}(\Delta t)$ in accord with the Born approximation,
\begin{align}
\label{eq:UT}
{U}_{\rm SE}^{[n]}(\Delta t)=1&-i H_{\rm S}(t_{n-1})\Delta t+\left(J_{\rm S}(t_{n-1}) \Delta B^{\dagger }_{[n]}-{\rm h.c.}\right)\nonumber\\
&-\frac{1 }{2}J_{\rm S}(t_{n-1})J_{\rm S}^{\dagger }(t_{n-1})\Delta B^{\dagger }_{[n]} \Delta B_{[n]}\nonumber\\
&-\frac{1}{2}J_{\rm S}^{\dagger }(t_{n-1})J_{\rm S}(t_{n-1})\Delta B_{[n]}\Delta B^{\dagger }_{[n]}\nonumber\\
&+\frac{1}{2}\left(J^2_{\rm S}(t_{n-1})\Delta B^{\dagger 2}_{[n]}+{\rm h.c.}\right).
\end{align}
Physically, Eq.~\eqref{eq:UT} describes the exchange of excitations between the system and the $n$th time-bin mode.

The overall time evolution operator in the interaction picture thus amounts to the sequential interaction between the system and each time bin $[n]$, with $n=1,2,\dots N$, and can be expressed as 
\begin{equation}
\label{eq: evolution_int}
U_{\rm SE}(T)={\cal T}\prod_{n=1}^N U_{{\rm SE}}^{[n]}(\Delta t)
\end{equation}
with ${\cal T}$ denoting the time-ordering. Such a sequential interaction determines a natural MPS structure~\cite{PhysRevLett.95.110503,PhysRevLett.105.260401,PhysRevLett.116.093601} of the state of the emission field, as we discuss below.

\subsection{Matrix Product State Representation}
We assume that at the initial time $t=0$ the system and the environment are decoupled and, as is typical in quantum optics, that the environment is initially in the vacuum state~\footnote{This formulation also applies to scenarios with coherent input fields, which is equivalent to the vacuum input plus classical driving fields; and can be straightforwardly extended to scenarios with arbitrary pure Gaussian input fields, see, e.g., Refs.~\cite{Wiseman,gardiner2015quantum}} $|{\rm vac}\rangle\equiv\otimes_{n=1}^\infty \ket{0}_{n}$, with $\ket{0}_{n}$ the Fock vacuum of the time bin $[n]$. The initial global state can therefore be expressed as $\ket{\Psi(0)}=\ket{\psi_{\rm S}(0)}\otimes\ket{\rm vac}$, with $\ket{\psi_{\rm S}(0)}$ the initial pure state of the system. 

The time-evolved global state $\ket{\Psi(T)}=U_{\rm SE}(T)\ket{\Psi(0)}$, with $U_{\rm SE}(T)$ defined in Eq.~\eqref{eq: evolution_int}, can be expanded in the time-bin Fock basis $\ket{\sigma}_{n}$, $\sigma_n=0,1$ as~\cite{PhysRevLett.95.110503,PhysRevLett.105.260401,PhysRevLett.116.093601} 
\begin{equation}
\label{eq:MPSglobal}
\ket{\Psi(T)}=\sum_{\{\sigma_n\}} A_{[N]}^{\sigma_N}\dots A_{[2]}^{\sigma_2}A_{[1]}^{\sigma_1}\ket{\psi_{\rm S}(0)}\otimes\ket{\sigma_N\dots\sigma_2\sigma_1}.
\end{equation}
Here, $A_{[n]}^{\sigma_n}$ are Kraus operators of the system. Assuming the dimension of the system Hilbert space is $D$, $A_{[n]}^{\sigma_n}$ are $D\times D$ matrices defined by expressing the elementary evolution operator ${U}_{\rm SE}^{[n]}(\Delta t)$ in the Fock basis of the time-bin, $A_{[n]}^{\sigma_n}\equiv\langle \sigma_n | U_{\rm SE}^{[n]}(\Delta t)|0_n\rangle$, $\sigma_n=0,1$. Using Eq.~\eqref{eq:UT}, the result can be expressed explicitly in terms of the system Hamiltonian and jump operator as
\begin{align}
\label{eq:KrausOP}
A_{[n]}^0&=1-i H_{\rm S}(t_{n-1})\Delta t-\frac{1}{2} J_{\rm S}^{\dagger }(t_{n-1})J_{\rm S}(t_{n-1}) \Delta t,\nonumber\\
A_{[n]}^1&=\sqrt{ \Delta t }J_S(t_{n-1}).
\end{align}

Equation~\eqref{eq:MPSglobal} is an MPS, featuring (finite-range) temporal correlations between the individual time bins $[n]$. The strength of the correlations depends on the bond dimension, i.e., the dimension of the constituting matrices $A_{[n]}^{\sigma_n}$, which is identical to the system Hilbert space dimension $D$. 
Such correlations originate from the quantum evolution according to Eq.~\eqref{eq: evolution_int}, which correlates the system and the time bins $[n]$, $n=1,2,\dots N$ sequentially via the dissipative interaction, therefor establishing correlations among the time bins. The nature of such sequential evolution is inherited by the structure of the state Eq.~\eqref{eq:MPSglobal}, of which each term in the summation $\sum_{\{\sigma_n\}}$ is generated by the sequential application of a series of Kraus operators $A_{[N]}^{\sigma_N}\dots A_{[2]}^{\sigma_2}A_{[1]}^{\sigma_1}$ to the system state $\ket{\psi_{\rm S}(0)}$, conditioned on a specific environment state $\ket{\sigma_N\dots\sigma_2\sigma_1}$.

We further define the reduced state of the environment by tracing out the system, $\rho_{\rm E}(T)={\rm tr_S}(\ket{\Psi(T)}\bra{\Psi(T)})$. Based on Eq.~\eqref{eq:MPSglobal}, it can be represented as a matrix product density operator (MPDO)
\begin{align}
\label{eq:MPDOenv}
\rho_{\rm E}(T)=\sum_{\{\sigma_n,\sigma_n^\prime\}}&{\rm tr}\left({\cal A}^{\sigma_N,\sigma_N^\prime}_{[N]}\dots{\cal A}^{\sigma_2,\sigma_2^\prime}_{[2]}{\cal A}^{\sigma_1,\sigma_1^\prime}_{[1]}\rho_{\rm S}(0)\right)\nonumber\\
&\ket{\sigma_N\dots\sigma_2\sigma_1}\bra{\sigma_N^\prime\dots\sigma_2^\prime\sigma_1^\prime},
\end{align}
where ${\cal A}^{\sigma_n,\sigma_n^\prime}_{[n]}$ are superoperators of the system defined via ${\cal A}^{\sigma_n,\sigma_n^\prime}_{[n]}(\cdot):= A_{[n]}^{\sigma_n}(\cdot)A_{[n]}^{\sigma_n^\prime\dag}$, and $\rho_{\rm S}(0)=\ket{\psi_{\rm S}(0)}\bra{\psi_{\rm S}(0)}$ is the initial system density operator. As such, $\rho_{\rm E}(T)$ inherits the structure of correlations among the time bins from the global state Eq.~\eqref{eq:MPSglobal}.

We comment that in Eqs.~\eqref{eq:MPSglobal} and \eqref{eq:MPDOenv} we have used the discrete MPS representation as a reflection of the finite time-bin width $\Delta t$ of the microscopic model, cf. Sec.~\ref{sec:coarse-graining}, rather than adopting the continuous MPS~\cite{PhysRevLett.104.190405} in the limit $\Delta t\to 0$. Nevertheless, for practical quantum-optical applications $\Delta t$ can be regarded as infinitesimally small such that the continuous limit can be safely and conveniently taken, as we illustrate in later discussions.

Finally, let us define the reduced state of the system by tracing out the environment, $\rho_{\rm S}(T)={\rm tr_E}(\ket{\Psi(T)}\bra{\Psi(T)})$. Using Eq.~\eqref{eq:MPSglobal}, it is straightforward to show that it obeys the update law $\rho_{\rm S}(T+\Delta t)=\sum_{\sigma_{N+1}}A_{[N+1]}^{\sigma_{N+1}}\rho_{\rm S}(T)A_{[N+1]}^{\sigma_{N+1}\dag}$. Taking the continuous limit $\Delta t\to 0$, this recovers the familiar Lindblad master equation (LME) for the reduced dynamics of the open system,
\begin{equation}
\label{eq:ME}
\dot{\rho}_{\rm S}(T)=-i[H_{\rm S}(T),\rho_{\rm S}]+{\cal D}[J_{\rm S}(T)]\rho_{\rm S},
\end{equation}
where we have defined the Lindblad operator ${\cal D}[J]\rho\equiv J\rho J^\dag-\frac{1}{2}\{J^\dag J,\rho\}$.

In summary, driven-dissipative quantum optical systems emits correlated photonic field as captured by the MPSs. The elementary building blocks of these states are the individual tensors $A_{[n]}$, which depend on the dynamic parameters of the system via Eq.~\eqref{eq:KrausOP}. As we discuss in the following, this allows for the sensing of unknown parameters via continuous detection of the emission field.

\section{Quantum Sensing via continuous measurement of the emission field}
\label{sec:eQFI}
We now introduce quantum sensing via continuous measurement, based on the MPS framework introduced above. Without loss of generality, we consider the sensing of a single quantity $\theta$,  which we assume is encoded in the system dynamics and thus the tensors $A_{[n]}(\theta)$. We focus on the `neighborhood' sensing scenario, cf. Fig.~\ref{fig:fig1}, where our task is to precisely determine the true value of the unknown parameter, $\theta=\theta_0+\delta$, around a rough estimator $\theta_0$ that is already known from preliminary information, as in typical sensing experiments~\cite{RevModPhys.87.637,RevModPhys.89.035002}. Our primary interest  in the following is to construct an efficient measurement strategy to reach the QCRB in the neighborhood of $\theta_0$. To this end, we will first introduce the achievable precision by continuous measurement, as quantified by the (classical) Fisher information (FI). We will then discuss the optimal precision that we aim to achieve, the QFI of the emission field.

\subsection{Fisher Information of Continuous Measurement}
\label{eq:FI_counting}
Let us consider measurement of the emission field, including counting and homodyning as familiar examples. Theoretically, such measurement can be represented by a series of positive operator-valued measures (POVMs) $\Pi_{D(0,T)}$ associated with the (continuous) measurement records $D(0,T)$ spanning the time interval $[0,T]$. The probability distribution of the measurement records, $P_{\theta}[D(0,T)]:= {\rm tr}[\rho_{\rm E}(\theta,T) \Pi_{D(0,T)}]$, is dependent on $\theta$, allowing us to estimate $\theta$ by processing of the measurement records. The associated precision can be quantified by the Fisher information (FI)
\begin{equation}
\label{eq:FI}
F(\theta, T)=\sum _{D(0,T)} P_{\theta}[D(0,T)]\left\{\partial _{\theta }\ln  P_\theta[D(0,T)]\right\}{}^2.
\end{equation}
According to the Cram\'er-Rao inequality~\cite{Kay97}, $F(\theta_0, T)$ sets a lower bound to the variance of any (unbiased) estimator $\theta_{\rm est}$ around $\theta_0$, i.e., 
${\rm Var}(\theta_{\rm est})\geq 1/{ K F}(\theta_0,T)$, where $K$ is the number of sensing interrogations.

As a familiar example, photon counting corresponds to projective measurement in the product of time-bin Fock basis as introduced in Sec.~\ref{sec:minimalModel}, with the counting records signalling the presence or absence of photons in each time bin, $D(0,T)=\{\sigma_n\}\equiv\{\sigma_1,\sigma_2,\dots,\sigma_N\}$, $\sigma_n=0,1$. The POVMs for photon counting are therefore
\begin{equation}
\label{eq:countingPOVM}
\Pi_{\{\sigma_n\}}=\otimes_{n=1}^{N} |\sigma\rangle_n\langle\sigma|.
\end{equation}
The probability distribution can be calculated from Eq.~\eqref{eq:MPDOenv} as $P_{\theta}[D(0,T)=\{\sigma_n\}]={\rm tr}[\tilde{\rho}_{{\rm S},c}(T)]$, where $\tilde{\rho}_{{\rm S},c}(T)={\cal T}\prod_{n=1}^N{\cal A}^{\sigma_n,\sigma_n}_{[n]}(\theta)\rho_{\rm S}(0)$ is an (unnormalized)~\footnote{Throughout this article, the upper tilde is used to indicate an unnormalized quantum state} conditional density matrix of the sensor that can be propagated efficiently via quantum trajectory simulation~\cite{carmichael1993open,RevModPhys.70.101,gardiner2004quantum,Wiseman,gardiner2015quantum}. Statistical average over sufficient numbers of trajectories allows for numerical evaluation of the FI Eq.~\eqref{eq:FI}. Homo(Hetero)-dyne measurement and the associated FI can be formulated similarly. 

A common feature of these familiar continuous measurements is that they are temporally local, i.e., their POVMs are separable in the time-bin basis. These measurements, however, may not be capable to extract the full information of the parameter, as quantified by the QFI of the emission field that we introduce below.

\subsection{Quantum Fisher Information of the Emission Field}
\label{QFIe}
According to the quantum Cram\'er-Rao inequality~\cite{PhysRevLett.72.3439}, the QFI of the emission field sets an upper limit to the estimation precision achieved by any measurement strategy of the emission field, $F(\theta_0,T) \leq I_{\rm E}(\theta_0,T)$. As a result, the QCRB of our sensing scenario can be expressed as $1/K I_{\rm E}(\theta_0,T)$.

The QFI captures the sensitivity of the environmental state Eq.~\eqref{eq:MPDOenv} to a small variation of the unknown parameter, $\theta\to\theta+\delta$,
\begin{equation}
\label{eq:QFIe}
I_{\rm E}(\theta,T)=-4\partial_\delta^2 {\cal F}_{\rm E}(\theta,\theta+\delta)|_{\delta=0}.
\end{equation}
where ${\cal F}_{\rm E}(\theta_1,\theta_2)$ is the overlap between the environmental state Eq.~\eqref{eq:MPDOenv} at two values $\theta_{1,2}$ of the unknown parameter, as quantified by the quantum fidelity
\begin{equation}
\label{eq:quantum_fidelity_env}
{\cal F}_{\rm E}(\theta_1,\theta_2)= {\rm tr}\left(\sqrt{\sqrt{\rho_{\rm E}(\theta_1,T)}\rho_{\rm E}(\theta_2,T) \sqrt{\rho_{\rm E}(\theta_1,T)}}\right).
\end{equation}

Despite its fundamental importance, the evaluation of $I_{\rm E}(\theta,T)$ for generic sensor designs remains an outstanding open challenge due to the complex structure of the environmental state, cf. Eq.~\eqref{eq:MPDOenv}. As a first significant result, we overcome this challenge by deriving a simple analytical formula for the evaluation of ${\cal F}_{\rm E}(\theta_1,\theta_2)$ and thus $I_{\rm E}(\theta,T)$, which only requires the knowledge of the sensor evolution, i.e., the LME~\eqref{eq:ME}. To keep our presentation concise, we summarize our main result here, and defer a full derivation to Appendix~\ref{eq: QFIe}. The quantum fidelity can be determined via
\begin{equation}
\label{eq: envQFI_fomula}
{\cal F}_{\rm E}(\theta_1,\theta_2)={\rm tr}\left[\sqrt{\mu_{\theta_1,\theta_2}(T)\mu_{\theta_1,\theta_2}^\dag(T)}\right],
\end{equation}
where $\mu_{\theta_1,\theta_2}(T)$ is a generalized density operator of the open sensor, as first introduced in Ref.~\cite{PhysRevLett.112.170401}
\begin{equation}
\label{eq:generalizedDM}
\mu_{\theta_1,\theta_2}(T)={\rm tr}_{\rm E}(|\Psi(\theta_1,T)\rangle\langle \Psi(\theta_2,T)|).
\end{equation}
Exploiting Eq.~\eqref{eq:MPSglobal}, it is straightforward to show that $\mu_{\theta_1,\theta_2}(T)$ satisfies the initial condition $\mu_{\theta_1,\theta_2}(0)=\rho_{\rm S}(0)$, and the update law 
\begin{equation}
\mu_{\theta_1,\theta_2}(T+\Delta t)=\sum_{\sigma_{N+1}}A_{[N+1]}^{\sigma_{N+1}}(\theta_1)\mu_{\theta_1,\theta_2}(T)A_{[N+1]}^{\sigma_{N+1}\dag}(\theta_2)\nonumber
\end{equation}
which can be expressed in the limit $\Delta t\to 0$ as a differential equation~\cite{PhysRevLett.112.170401} [we denote $\mu(t)\equiv\mu_{\theta_1,\theta_2}(t)$ below for notational simplicity],
\begin{align}
\label{eq:generalizedME}
\frac{d\mu}{dt}=&-i\left[H_{\rm S}(\theta_1,t)\mu-\mu H_{\rm S}^\dag(\theta_2,t)\right]+J_{\rm S}(\theta_1,t)\mu J_{\rm S}^\dag(\theta_2,t)\nonumber\\
&-\frac{1}{2}\left[J_{\rm S}^\dag(\theta_1,t) J_{\rm S}(\theta_1,t) \mu + \mu J_{\rm S}^\dag(\theta_2,t) J_{\rm S}(\theta_2,t)\right]\nonumber\\
:=& \,{\cal L}(\theta_1,\theta_2,t)\,\mu,
\end{align}
and can be propagated efficiently for sensor designs with a modest Hilbert space dimension.

Equation~\eqref{eq:generalizedME} describes, in general, a quantum map that is not completely positive trace-preserving (CPTP). It reduces to the LME \eqref{eq:ME}, thus recovering CPTP, by setting $\theta_1=\theta_2=\theta$. In the neighborhood of $(\theta_1,\theta_2)=(\theta,\theta)$, the spectrum of ${\cal L}(\theta_1,\theta_2,t)$ can be smoothly connected to the spectrum of the LME~\eqref{eq:ME}. For many important sensor designs, the assoicated LME \eqref{eq:ME} is time homogeneous and ergodic (i.e., supports a unique stationary state), ${H}_{\rm S}(\theta,t)={H}_{\rm S}(\theta),{J}_{\rm S}(\theta,t)={J}_{\rm S}(\theta)$ and ${\rho}_{\rm S}(\theta,t\to\infty)=\rho_{\rm S}^{\rm st}(\theta)$. Correspondingly, ${\cal L}(\theta_1,\theta_2)$ is time independent, and has a unique eigenvalue $\lambda(\theta_1,\theta_2)$ that is smoothly connected to the vanishing eigenvalue of the LME \eqref{eq:ME}, ${\cal \lambda}(\theta,\theta)=0$; whereas the rest eigenvalues have a finite negative real part. The (right) eigenstate $\mu_{\theta_1,\theta_2}^\lambda$ associated with $\lambda(\theta_1,\theta_2)$ is smoothly connected to the sensor stationary state, $\mu_{\theta,\theta}^\lambda=\rho_{\rm S}^{\rm st}(\theta)$. We thus have, for long interrogation time $T$,
\begin{equation}
\label{mu:ergodic}
\mu_{\theta_1,\theta_2}(T)\sim \mu_{\theta_1,\theta_2}^\lambda {\exp}[-\lambda(\theta_1,\theta_2)T]
\end{equation}
and consequently ${\cal F}_{\rm E}(\theta_1,\theta_2)\sim {\rm exp}[-{\rm Re}\lambda(\theta_1,\theta_2)T]$, with Re denoting the real part. This exponentially decaying fidelity, via Eq.~\eqref{eq:QFIe}, dictates a linear scaling of the QFI of the emission field with respect to the interrogation time,
\begin{equation}
I_{\rm E}(\theta,T)\sim4T\partial_\delta^2{\rm Re}\lambda(\theta,\theta+\delta)|_{\delta=0}.
\end{equation}

In contrast, if the sensor evolution is not ergodic, the QFI of the emission field does not necessarily obey a linear scaling~\cite{PhysRevLett.112.170401,PhysRevA.93.022103}. From the point of view of MPS, if a driven-dissipative system is ergodic, its emission field is finitely correlated and has an \emph{injective} MPS representation~\cite{RevModPhys.93.045003}; if non-ergodic, the emission field may be temporally long-range correlated, lacking an injective MPS representation. In Sec.~\ref{sec:examples}, we will study a few models of open sensors, including both ergodic and non-ergodic ones, for which the relevant QFIs demonstrate drastically different behavior.
 
The effective method established in this section for the calculation of the environmental QFI can be extended straightforwardly to sensors coupled with multiple environments, which we detail In Appendix \ref{app:A3}.

\subsection{Global Quantum Fisher Information}
\label{sec:QFIg}
We complete this section with a brief discussion of a closely related precision bound, the global QFI $I_{\rm G}(\theta,T)$ as introduced in Refs.~\cite{PhysRevLett.106.090401,PhysRevLett.112.170401,2015JPhA...48J5301C}, and its relation with $I_{\rm E}(\theta,T)$ established above. The global QFI measures the sensitivity of the global state Eq.~\eqref{eq:MPSglobal} to a small variation of the unknown parameter,
\begin{align}
\label{eq:QFIg}
I_{\rm G}(\theta,T)=-4\partial_\delta^2 {\cal F}_{\rm G}(\theta,\theta+\delta)|_{\delta=0},
\end{align}
where ${\cal F}_{\rm G}(\theta_1,\theta_2)=|\langle\Psi(\theta_1,T)|\Psi(\theta_2,T)\rangle|$ is the quantum fidelity of the global (pure) state. This fidelity can be conveniently evaluated via the generalized density operator Eq.~\eqref{eq:generalizedDM} as~\cite{PhysRevLett.112.170401}
\begin{equation}
\label{eq:FG}
{\cal F}_{\rm G}(\theta_1,\theta_2)=|{\rm tr_S}\left[\mu_{\theta_1,\theta_2}(T)\right]|.
\end{equation}

As the global state involves both the sensor and the emission field, in general $I_{\rm G}(\theta,T)\geq I_{\rm E}(\theta,T)$. Moreover, the retrieval of $I_{\rm G}(\theta,T)$ may require (potentially experimentally challenging) joint measurement of the sensor and the emission field. If the sensor LME \eqref{eq:ME} is time-homogeneous and ergodic, then Eq.~\eqref{mu:ergodic} dictates, for long interrogation time $T$,
\begin{equation}
I_{\rm G}(\theta,T)\sim4T\partial_\delta^2{\rm Re}\lambda(\theta,\theta+\delta)|_{\delta=0}.
\end{equation}
Thus, $I_{\rm G}(\theta,T)$ exceeds $I_{\rm E}(\theta,T)$ by an (asymptotically) time-independent constant, and obeys the same long-time scaling. In contrast, if the evolution of the open sensor is non-ergodic, the difference between the two precision bounds may grow with time and consequently, leading to different asymptotic scaling in the long time limit. Such connection and difference will be illustrated in Sec.~\ref{sec:examples} in terms of paradigmatic models of open sensors.

\section{Efficient retrieval of the quantum Fisher Information}
\label{eq:eff_retrieval}
Our aim below is to devise an effective measurement scheme for the emission field of generic driven-dissipative sensors, to retrieve efficiently the emission-field QFI (up to a time-independent constant) thus to saturate the CQRB $1/K I_{\rm E}(\theta_0,T)$ for long interrogations. 
While we focus below on the continuous measurement of the emission field alone, we note as a prospect that a final strong measurement of the sensor subsequent to the continuous monitoring may provide additional useful information and may allow for the retrieval of the global QFI of the sensor and the emission field~\cite{Albarelli2018restoringheisenberg,Albarelli_2017}.

As outlined in the Sec.~\ref{sec:intro}, our scheme features the implementation of a temporally quasi-local continuous measurement, enabled by directing the emission field of the sensor into a quantum decoder module---an auxiliary open system subjected to continuous measurement at its output port, cf. Fig.~\ref{fig:fig1}. Our measurement scheme thus implements continuous monitoring of cascaded quantum optical setups~\cite{PhysRevLett.70.2269,PhysRevLett.70.2273}. 
The dynamics associated with the quantum decoder can be described by Eqs.~(\ref{eq:Hse}--\ref{eq: evolution_int}) with the simple replacement ${\rm S\to D}$ of the subscripts. In particular, the joint evolution operator of the environment and the decoder $U_{\rm DE}(T)$ can be transcribed from Eqs.~\eqref{eq:UT} and \eqref{eq: evolution_int}, and features the sequential interaction with the time bins $[n]$, $n=1,2,\dots,N$. We emphasize that the decoder interacts with the time bins \emph{subsequent} to the sensor thus guaranteeing causality, as naturally incorporated in the theory of cascaded open systems~\cite{PhysRevLett.70.2269,PhysRevLett.70.2273}.

To be specific, we consider below continuous counting of the output field of the quantum decoder, as captured by the projectors $\Pi_{\{\sigma_n\}}$ in Eq.~\eqref{eq:countingPOVM}. This effectively implements a measurement 
\begin{equation}
\label{eq:MPS_meas}
\Pi^{\rm DE}_{\{\sigma_n\}}=U_{\rm DE}^\dag(T)\Pi_{\{\sigma_n\}}U_{\rm DE}(T)
\end{equation}
of the decoder-environment state $\rho_{\rm D}(0)\otimes\rho_{\rm E}(T)$ prior to their interaction, with $\rho_{\rm D}(0)=|\psi_{\rm D}(0)\rangle\langle \psi_{\rm D}(0)|$ being the initial state of the decoder, and $\rho_{\rm E}(T)$ being exactly the emission field state we wish to measure, cf. Eq.~\eqref{eq:MPDOenv}. The effective measurement Eq.~\eqref{eq:MPS_meas} is distinguished from conventional continuous measurement by two unique features. First, the sequential character of the evolution operator $U_{\rm DE}(T)$ renders $\Pi^{\rm DE}_{\{\sigma_n\}}$ a natural matrix-product structure, representing temporally finitely-correlated continuous measurement. Secondly, such a measurement can be efficiently engineered by controlling the evolution operator of the decoder $U_{\rm DE}(T)$.

In the following, we will harness such a controllability to implement an optimal effective measurement $\Pi^{\rm DE}_{\{\sigma_n\}}$ for efficiently retrieving the QFI of the emission field. To this end, let us introduce the condition for an optimal measurement, and convert it to a requirement to the evolution operator of the decoder $U_{\rm DE}(T)$.

\subsection{Optimal Measurement of the Emission Field}
\label{sec:optimal_meas}
The optimal effective measurement of the emission field can be revealed by analyzing the structure of the global state $\ket{\Psi(\theta,T)}$, cf.~Eq.~\eqref{eq:MPSglobal}. As $\ket{\Psi(\theta,T)}$ is a pure state, the optimal measurement of $\ket{\Psi(\theta,T)}$, capable of retrieving the global QFI in the neighborhood of $\theta_0$, is conveniently provided~\cite{PhysRevLett.72.3439}. In particular, any projective measurement $\{\Pi_{k}^{\rm SE}\}$~\footnote{For a projective measurement $\{\Pi_{k}^{\rm SE}\}$, the projectors have to be mutually orthogonal, i.e., $\Pi_{k}^{\rm SE} \Pi_{k'}^{\rm SE}=\delta_{k,k'}\Pi_{k}^{\rm SE}$.} of $\ket{\Psi(\theta,T)}$ involving a specific projector
\begin{align}
\label{eq:optimal_basis}
\Pi_{0}^{\rm SE}=|\Psi (\theta_0, T)\rangle\langle \Psi (\theta_0, T)|
\end{align}
is optimal~\cite{smerzi2,PhysRevLett.119.130504,Liu_2020}. Physically, the optimal precision is rooted in the fact that small variations of the parameter, $\theta=\theta_0+\delta$, introduces sizeable change to the probability distribution $P_{\theta}(\Pi_k^{\rm SE})$ associated with the projectors $\Pi_{k}^{\rm SE},k\neq 0$~\cite{smerzi2,PhysRevLett.119.130504,Liu_2020}. Note that such a projective measurement is generally ineffective in identifying the sign of the small variation $\delta$, as $P_{\theta_0\pm\delta}(\Pi_k^{\rm SE})$ is symmetric with respect to $\theta_0$ for small enough $\delta$~\cite{madalin}. Such unidentifiability can nevertheless be circumvented in practice by deliberately choosing $\theta_0$ such that it is outside (but sufficiently close to) the confidence interval of a reasonably good prior estimator~\cite{madalin}, see Appendix~\ref{app:B5} for a brief discussion.

We emphasize that Eq.~\eqref{eq:optimal_basis} represents a correlated joint measurement of the sensor and the emission field (environment), thus may be challenging to implement in practice. Nevertheless, it naturally possesses a matrix-product structure regarding the environmental DOFs, i.e., the time bins, as inherited from the global state $|\Psi (\theta_0, T)\rangle$. Remarkably, such a matrix-product structure also appears naturally in our effective measurement $\{\Pi^{\rm DE}_{\{\sigma_n\}}\}$ enabled by the quantum decoder, cf. Eq.~\eqref{eq:MPS_meas}.

To optimize the effective measurement Eq.~\eqref{eq:MPS_meas}, we thus require the projectors $\Pi^{\rm DE}_{\{\sigma_n\}}$ to have the same structure as $\Pi^{\rm SE}_k$ concerning the environmental DOFs.
As $\Pi^{\rm SE}_{k\neq 0}$ are essentially arbitrary, this narrows down to requiring, for a specific counting signal $\{\sigma_n^*\}$, that $\Pi^{\rm DE}_{\{\sigma_n^*\}}$ has the same structure as  $\Pi^{\rm SE}_0$ concerning the environment.
Such a condition can be satisfied by enforcing 

\begin{equation}
\label{eq:optimal_cond}
U_{\rm DE}(\theta_0,T)\left[\Pi_{0}^{\rm SE}\otimes \rho_{\rm D}(0) \right]U_{\rm DE}^\dag(\theta_0,T)=\Pi^{\rm}_{\{\sigma_n^*\}}\otimes\rho_{\rm SD}(\theta_0,T),
\end{equation}
i.e., the quantum evolution involving the decoder, $U_{\rm DE}(\theta_0,T)$, decodes the environmental matrix-product structure in $\Pi_{0}^{\rm SE}$ into a product structure, as captured by the (standard photon-counting) projector $\Pi^{\rm}_{\{\sigma_n^*\}}$ corresponding to the signal $\{\sigma_n^*\}$, c.f. Eq.~\eqref{eq:countingPOVM}. Such evolution leaves the remaining DOFs, i.e., the sensor and the decoder, in a general state $\rho_{\rm SD}(\theta_0,T)$ which is not measured. Note that we have denoted explicitly the dependence of $U_{\rm DE}$ and $\rho_{\rm SD}$ in Eq.~\eqref{eq:optimal_cond} on the prior information $\theta_0$, as inherited from Eq.~\eqref{eq:optimal_basis}.

The choice of the counting signal $\{\sigma_n^*\}$ in Eq.~\eqref{eq:optimal_cond} is arbitrary. For concreteness, we will adopt the choice $\{\sigma_n^*\}=\{0,0,\dots, 0\}$, i.e., the detection of null photons, and correspondingly
 \begin{equation}
 \label{eq:vac_projector}
 \Pi_{\{\sigma_n^*\}}=\otimes_{n=1}^{N} |0\rangle_n\langle0|, 
 \end{equation}
a projector to the vacuum state of time bins $n\in[1,N]$. 

To summarize, we have expressed the optimal measurement of the emission field in terms of an \emph{a priori} condition to the evolution of the decoder, as captured by Eqs.~\eqref{eq:optimal_cond} and \eqref{eq:vac_projector}. While we have constructed such a condition on the basis of physical argument, we rigorously prove its long-time optimum in Appendix~\ref{app:optimal_proof}, by showing that for any sensor whose evolution is ergodic, the measurement satisfying Eqs.~\eqref{eq:optimal_cond} and \eqref{eq:vac_projector} allows for the retrieval of the QFI of the emission field up to a finite, time-independent constant. In the next section we will show that requirement \eqref{eq:optimal_cond} and \eqref{eq:vac_projector} can indeed be satisfied via an appropriate implementation of the decoder, i.e., via a proper choice of its initial state $|\psi_{\rm D}(0)\rangle$ and evolution $U_{\rm DE}(\theta_0,T)$, and that we can implement them efficiently.

\subsection{Implementation of the Optimal Measurement via the Quantum Decoder}
\label{sec:decoder_recipe}
\begin{figure*}[t!]
\centering{} \includegraphics[width=1\textwidth]{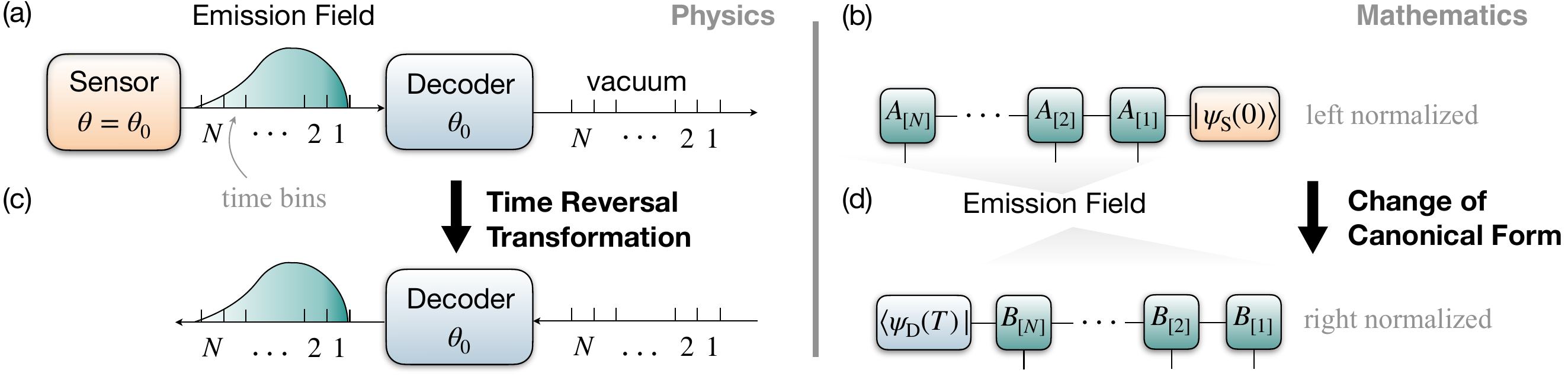} 
\caption{Illustration of the decoder design in terms of time reversal transformation in quantum optical input-output channels. (a) The sensor interacts with the time bins $[n],n=1,2,...,N$ sequentially, resulting in the sensor-environment global state $\ket{\Psi({\theta},T)}$, cf. Eq.~\eqref{eq:MPSglobal}, whose tensor diagram is shown in (b) under the simplifying assumption that the sensor and environment disentangle at time $t=T$. The decoder interacts with the time bins subsequent to the sensor. (b) The tensors $A_{[n]}(\theta)$, cf. Eq.~\eqref{eq:KrausOP}, satisfy the left normalization condition $\sum_{\sigma_n}A_{[n]}^{\sigma_n\dag}(\theta) A_{[n]}^{\sigma_n}(\theta)=\mathbb{I}_{D}$. (c) The optimal measurement condition Eq.~\eqref{eq:optimal_cond} requires, for $\theta=\theta_0$, that the joint unitary evolution of the decoder and the waveguide maps the sensor emission field to the vacuum. Consequently, time reversal of such evolution maps the environmental vacuum back to the sensor emission field. The time reversed evolution features sequential interaction between the decoder and the time bins $[n]$ in an anti-time-ordered manner, $n=N,N-1,\dots 1$. (d) The state generated by the anti-time-ordered interaction, for which the tensors $B_{[n]}$ satisfy the right normalization condition $\sum_{\sigma_n}B_{[n]}^{\sigma_n} B_{[n]}^{\sigma_n\dag}=\mathbb{I}_{D}$. By requiring that this state matches the sensor emission field at $\theta=\theta_0$, $B_{[n]}$ are related to $A_{[n]}(\theta_0)$ via the SVD procedure described in Sec.~\ref{sec:decoder_recipe}. While in this figure the sensor and the environment are assumed to disentangle at time $t=T$ to simplify the illustration, our general recipe for the decoder construction discussed in Sec.~\ref{sec:decoder_recipe} does not rely on this assumption and applies to general continuous measurement settings.}
\label{fig:fig1.5} 
\end{figure*}
Let us now construct an explicit recipe for achieving the optimal measurement Eq.~\eqref{eq:optimal_cond}, via engineering the evolution operator $U_{\rm DE}(\theta_0,T)$, or equivalently, the Hamiltonian $H_{\rm D}(\theta_0,t)$ and jump operator $J_{\rm D}(\theta_0,t)$ of the decoder. Such a construction is enabled by the following theorem.

\emph{
\textbf{Theorem.} Equation~\eqref{eq:optimal_cond} can be fulfilled by choosing the decoder to be a driven-dissipative quantum system that has the same Hilbert space as the sensor, with an initial state $|\psi_{\rm D}(0)\rangle=W_0^\dag|\psi_{\rm S}(0)\rangle$, where $W_0$ is an arbitrary $D\times D$ unitary rotation, and the Hamiltonian $H_{\rm D}(\theta_0,t)$ and the jump operator $J_{\rm D}(\theta_0,t)$ of the decoder to fulfill
\begin{align}
\bar{B}_{[n]}^0(\theta_0)=&\,1-i H_{\rm D}(\theta_0,t_{n-1})\Delta t\nonumber\\
&-\frac{1}{2}J_{\rm D}^{\dagger }(\theta_0,t_{n-1})J_{\rm D}(\theta_0,t_{n-1})\Delta t,\nonumber\\
\bar{B}_{[n]}^1(\theta_0)=&-\sqrt{\Delta t}J_{\rm D}^{\dagger }(\theta_0,t_{n-1}).
\label{eq:Kraus_anti}
\end{align}
Here $\bar{B}_{[n]}(\theta_0)$ denotes the complex conjugate of the tensor $B_{[n]}(\theta_0)$, which is provided via right normalization of the MPS $\ket{\Psi(\theta_0,T)}$. This can be done by sequential application of singular value decompositions (SVDs). We start by writing $A_{[1]}(\theta_0)(|\psi_{\rm S}(0)\rangle\langle\psi_{\rm D}(0)|\otimes \mathbb{I}_{2})=R_{[1]}(\theta_0) B_{[1]}(\theta_0)$, where $\mathbb{I}_2$ is the ${2\times2}$ identity matrix, $B_{[1]}(\theta_0)$ is the right unitary in the SVD and $R_{[1]}(\theta_0)$ is the remaining part. Continuation of this procedure via the recipe
\begin{equation}
\label{eq:SVD_main}
A_{[n]}(\theta_0)[R_{[n-1]}(\theta_0)\otimes \mathbb{I}_2]=R_{[n]}(\theta_0)B_{[n]}(\theta_0)
\end{equation}
for the time bins $n=2,3,\dots N$, with $B_{[n]}(\theta_0)$ the right unitary in the SVD and $R_{[n]}(\theta_0)$ the remaining part, completes the right normalization.}

To keep our presentation concise, here we only provide a physical interpretation of the essence of the theorem, and defer its rigorous proof to Appendix \ref{app: proof_left_normalization}. The interpretation, as illustrated in Fig.~\ref{fig:fig1.5}, is based on generalizing the time-reversal argument underlying cascaded quantum state transfer~\cite{PhysRevLett.78.3221} to our present scenario, i.e., generalizing a single photon traversing the waveguide to (arbitrarily correlated) multiple photons: the optimal measurement condition, Eqs.~\eqref{eq:optimal_cond} and~\eqref{eq:vac_projector}, implies that $U_{\rm DE}(T)$ maps the input environmental state (that is, the sensor emission field) to the vacuum state for $\theta=\theta_0$. Time reversal of such evolution, therefore, maps the vacuum state back to the input state. Such time reversed evolution features sequential interaction between the decoder and the time bins $[n]$ in an anti-time-ordered manner, $n=N,N-1,\dots 1$, cf. Fig.~\ref{fig:fig1.5}, corresponding to changing the canonical form of the MPS.

The above SVD procedure can be recast as a differential equation in the limit $\Delta t\to 0$, allowing us to find the desired decoder evolution conveniently. To this end, let us define ${\rho}(t_n\equiv n \Delta t):=R_{[n]}(\theta_0)R_{[n]}^\dag(\theta_0)$, which satisfies the initial condition $\rho(0)=\ket{\psi_{\rm S}(0)}\bra{\psi_{\rm S}(0)}$. According to Eq.~\eqref{eq:SVD_main}, ${\rho}(t)$ obeys the update law ${\rho}(t)=\sum_{\sigma_n}A_{[n]}^{\sigma_n}(\theta_0){\rho}(t-\Delta t)A_{[n]}^{\sigma_n\dag}(\theta_0)$ which, in the limit $\Delta t\to 0$, is identical to the LME~\eqref{eq:ME} (with the unknown parameter set as $\theta_0$). Therefore, $\rho(t)$ is exactly the sensor density matrix, $\rho(t)\equiv\rho_{\rm S}(\theta_0,t)$. Given ${\rho}_{\rm S}(\theta_0,t)$, it is easy to show that the choice
\begin{equation}
\label{eq:R_general_solution}
R(\theta_0,t)=\sqrt{{\rho}_{\rm S} (\theta_0,t)} W(t)
\end{equation}
exhausts all legitimate solutions of the SVD Eq.~\eqref{eq:SVD_main}, where \(W(t)\) is an arbitrary time-dependent unitary matrix reflecting the gauge redundancy of the SVD, with the initial condition $W(0)=W_0$. Substituting Eqs.~\eqref{eq:KrausOP} and~\eqref{eq:Kraus_anti} into Eq.~\eqref{eq:SVD_main} provides us with explicit expressions of the Hamiltonian and the jump operator of the decoder,
\begin{align}
H_{\rm D}(\theta_0,t)=&-\frac{1}{2}\bigg\{R^{\rm T}(\theta_0,t)H_{\rm S, nh}^{\rm T}(\theta_0,t)\left[R^{\rm T}(\theta_0,t)\right]^{-1}\!+{\rm h.c.}\bigg\},\nonumber\\
J_{\rm D}(\theta_0,t)=&-R^{\rm T}(\theta_0,t){J}_{\rm S}^{\rm T}(\theta_0,t)\left[R^{\rm T}(\theta_0,t)\right]^{-1},
\label{eq:H_J_general_solution}
\end{align}
in which $H_{\rm S,nh}:=H_{\rm S}-i J_{\rm S}^\dag J_{\rm S}/2$ is a non-Hermitian Hamiltonian of the sensor, and the superscript T denotes matrix transpose. Equation~\eqref{eq:H_J_general_solution} serves as the central formula of this section, as it provides us with an explicit construction of the quantum decoder for the efficient retrieval of the QFI of the emission field, applicable to generic (including nonlinear and time dependent) open sensor designs. Remarkably, such a construction relies only on the knowledge of the LME \eqref{eq:ME} and its solution $\rho_{\rm S}(t)$.

As an example, let us apply Eq.~\eqref{eq:H_J_general_solution} to the special case of a time homogeneous LME supporting a unique stationary state, i.e., ${H}_{\rm S}(\theta_0,t)={H}_{\rm S}(\theta_0),{J}_{\rm S}(\theta_0,t)={J}_{\rm S}(\theta_0)$ and ${\rho}_{\rm S}(\theta_0,t\to\infty)=\rho_{\rm S}^{\rm st}(\theta_0)$. Using the spectrum decomposition $\rho_{\rm S}^{\rm st}(\theta_0)=\sum_{k=1}^D p_k^{\rm st}(\theta_0) \ket{k_{\rm st}(\theta_0)}\bra{k_{\rm st}(\theta_0)}$ and choosing $W(t)=W_0$ an arbitrary time-independent unitary, we can reduce Eq.~\eqref{eq:H_J_general_solution} to the time-stationary solution
\begin{align}
\label{eq:diss_dimer}
H_{\rm D}(\theta_0)=&-\frac{1}{2}\sum_{k,k'}\sqrt{\frac{p_{k'}^{\rm st}(\theta_0)}{p_{k}^{\rm st}(\theta_0)}}\bigg[\langle k_{\rm st}(\theta_0)|H_{\rm S,nh}(\theta_0)|k^\prime_{\rm st}(\theta_0)\rangle\nonumber\\
&\times|\tilde{k}^\prime_{\rm st}(\theta_0)\rangle\langle\tilde{k}_{\rm st}(\theta_0)|+{\rm h.c.}\bigg],\nonumber\\
J_{\rm D}(\theta_0)=&-\!\sum_{k,k'}\!\sqrt{\frac{p^{\rm st}_{k'}(\theta_0)}{p_k^{\rm st}(\theta_0)}}\langle k_{\rm st}(\theta_0)|J_{\rm S}
|k^\prime_{\rm st}(\theta_0)\rangle\nonumber\\
&\times|\tilde{k}^\prime_{\rm st}(\theta_0)\rangle\langle\tilde{k}_{\rm st}(\theta_0)|
\end{align}
in the $t\to\infty$ limit, where we have defined the rotated basis $\ket{\tilde{k}_{\rm st}(\theta_0)}=\sum_{k'}(W_0)_{k,k'}\ket{k^\prime_{\rm st}(\theta_0)}$. Equation~\eqref{eq:diss_dimer} was first established in Ref.~\cite{Stannigel_2012} as a construction of coherent quantum absorbers for engineering the stationary-state entanglement in cascaded quantum networks. Here, it naturally emerges as a solution of our general decoder implementation Eq.~\eqref{eq:H_J_general_solution} for the case of time-stationary open quantum dynamics. Remarkably, our construction goes significantly beyond the time-stationary scenario by exploiting the general time reversal transformation of quantum optical input-output channels. It therefore allows for the construction of the optimal decoder for generic (including time-dependent) open sensors, as we illustrate in Sec.~\ref{sec:examples}.

The decoder-assisted information retrieval scheme constructed above can be readily adapted to (closed) many-body lattice models that are described by discrete MPSs, as we detail in Appendix \ref{app:discreteMPS}.

\subsection{The Retrieved Fisher Information}
\label{sec: FI_decoding}
Let us now analyze the achievable Fisher information of our decoder-assisted measurement scheme. The scheme implements continuous monitoring of cascaded quantum optical setups, with the decoder evolving nontrivially according to Eq.~\eqref{eq:H_J_general_solution}. The dynamics of the sensor and the decoder, subjected to photon counting of the decoder output field, can be described as a (unnormalized) stochastic cascaded ME~\cite{PhysRevLett.70.2269,PhysRevLett.70.2273}
\begin{align}
\label{eq:SME}
d\tilde{\varrho_{\rm}}_{c}=&-i[H_{\rm S}(\theta, t)+H_{\rm D}(\theta_0,t)+H_{\rm casc}(t),\tilde{\varrho_{\rm}}_{c}] dt\nonumber\\
&-\frac{1}{2}\{J^\dag(t) J(t),\tilde{\varrho_{\rm}}_{c}\}dt+\left[dt J(t) \tilde{\varrho_{\rm}}_{c} J^\dag(t) -\tilde{\varrho_{\rm}}_{c} \right] d{\cal N}(t).
\end{align}
In Eq.~\eqref{eq:SME}, $J(t)=J_{\rm S}(\theta,t)+J_{\rm D}(\theta_0,t)$ and $H_{\text{casc}}(t)=\frac{i}{2}[J_{\rm D}(\theta_0,t) J_{\rm S}^{\dagger }(\theta,t)-J_{\rm S}(\theta,t) J_{\rm D}^{\dagger }(\theta_0,t)]$. The stochastic Poisson increment $d{\cal N}(t)$ can take two values: $d{\cal N}(t)=1$ with probability $p_1={\rm tr}(\tilde{\varrho}_c J^\dag J)dt/{\rm tr}(\tilde{\varrho}_c)$ and $d{\cal N}(t)=0$ with probability $p_0=1-p_1$. 

Corresponding to the conditional evolution Eq.~\eqref{eq:SME}, the photon counting signal up to time \(T=Ndt\) for a specific trajectory is \(D(0,T)=\{d{\cal N}(0),d{\cal N}(dt),\text{...},d{\cal N}(T)\}\). The
probability of this trajectory is
\begin{align}
\label{eq: prob_sme}
P_{\theta}[D(0,T)]=p_{d{\cal N}(0)}\cdots p_{d{\cal N}(T)}={\rm tr}[\tilde{\varrho}_c(T)].
\end{align}
By sampling enough number of quantum trajectories, we can extract the FI retrieved by our scheme via Eq.~\eqref{eq:FI}.

Equation~\eqref{eq:SME} can be extended straightforwardly to account for decoherence and imperfections in realistic experimental implementation, thus allowing us to examine the noise resilience of our information retrieval scheme, as we detail in Sec.~\ref{sec:imperfect}.

\subsection{The Sensor-Decoder Entanglement}
\label{sec:sensor-decoder-entanglement}
It has long been recognized that quantum-enhanced measurements are closely associated with the generation of entanglement. For example, in the `negative-mass' oscillator scheme, conditional Einstein-Poldosky-Rosen (EPR) entanglement is established between the sensor oscillator and the ancillary (negative-mass) oscillator~\cite{PhysRevLett.121.103602,PhysRevLett.102.020501}. In our information retrieval scheme, the evolution of the decoder is dependent non-trivially on the evolution of the sensor, leading to entanglement between them. Indeed, condition Eq.~\eqref{eq:optimal_cond} can be interpreted as entanglement swapping: it transforms the sensor-environment entanglement to the sensor-decoder entanglement by disentangling the environment with the rest for $\theta=\theta_0$. As a result, the sensor-decoder state $\rho_{\rm SD}(\theta_0,T)$, cf., Eq.~\eqref{eq:optimal_cond}, is a \emph{pure} entangled state in the absence of experimental imperfection and decoherence. As we show in detail in Appendix~\ref{app:B2}, it can be written as $\rho_{\rm SD}(\theta_0,T)=\ket{\psi_{\rm SD}(\theta_0,T)}\bra{\psi_{\rm SD}(\theta_0,T)}$, with $\ket{\psi_{\rm SD}(\theta_0,T)}$ a natural purification of the sensor density matrix $\rho_{\rm S}(\theta_0,T)$. Specifically, suppose the sensor density matrix at time $t=T$ can be spectrally decomposed as $\rho_{\rm S}{(\theta_0,T)}=\sum_{k=1}^D p_k(\theta_0,T) \ket{k(\theta_0,T)}\bra{k(\theta_0,T)}$, then
\begin{align}
\label{eq:SD_entanglement}
\ket{\psi_{\rm SD}(\theta_0,T)}=&\sum_{k=1}^D \sqrt{p_k(\theta_0,T)} \ket{k(\theta_0,T)}_{\rm S} \nonumber\\
&\otimes\left(W^\dag(T)\ket{k(\theta_0,T)}_{\rm D}\right),
\end{align}
where $W(T)$ is the time-dependent unitary (gauge) redundant matrix introduced in Sec.~\ref{sec:decoder_recipe}. For $\theta\neq\theta_0$, the entanglement swapping is not perfect: $\rho_{\rm SD}(\theta,T)$ is in a mixed entangled state due to residual entanglement with the environment.

{In the case that the sensor reaches its stationary state, $\rho_{\rm S}(\theta_0,T)=\rho^{\rm st}_{\rm S}(\theta_0)$, Eq.~\eqref{eq:SD_entanglement} is time independent and reduces to the purification of the stationary state of the sensor. Such stationary entangled state is dark, i.e., decoupled from the environment, allowing for engineering exotic quantum many-body phases exploiting dissipation, see Refs.~\cite{Stannigel_2012,PhysRevLett.113.237203,PhysRevA.91.042116}.}

\section{Protocols of Efficient Information Retrieval}
\label{sec:examples}
Let us proceed to illustrate our information retrieval scheme at the hand of paradigmatic models of open quantum sensors. We start with a concise illustration with linear quantum sensors. We then apply our scheme to nonlinear sensor designs as represented by driven-dissipative emitters. Finally, we demonstrate our scheme with a driven-dissipative many-body sensor consisting of a transverse-field Ising spin chain.

\subsection{Linear Quantum Sensors}
The linear sensor model captures the core ingredient of sensor technologies such as gravitational wave detectors, optomechanical force sensors and atomic gas magnetometers. In such a scenario, as we will show, the proposed quantum decoder reduces to a displaced `negative mass' oscillator~\cite{Moller:2017ta,PhysRevLett.102.020501,PhysRevLett.105.123601,PhysRevLett.121.103602,PhysRevLett.121.031101}.

A linear quantum sensor is a continuous variable system with a pair of conjugated quadratures $X_{\rm S}$ and $P_{\rm S}$ obeying the canonical commutation relation $[X_{\rm S},P_{\rm S}]=i$. These can, e.g., be the position and momentum operators of a mechanical oscillator, or be the collective internal spins along two orthogonal directions of an atomic gas. The unknown parameter (e.g., a weak force) $f$ couples with $X_{\rm S}$, resulting in the sensor Hamiltonian
\begin{equation}
H_{\rm S} = \frac{1}{2}\omega (X_{\rm S}^2 + P_{\rm S}^2)-f X_{\rm S},
\end{equation}
with $\omega$ being the oscillation frequency. 

\begin{figure}[t!]
\centering{} \includegraphics[width=0.48\textwidth]{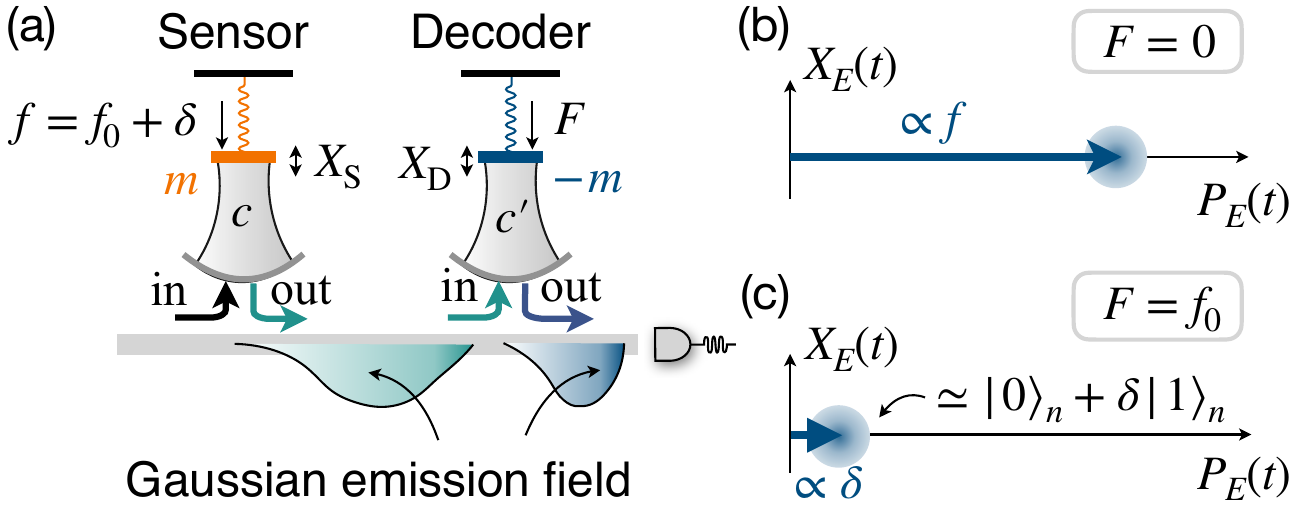} 
\caption{Efficient information retrieval for a linear optomechanical force sensor. (a) The unknown force $f$ couples with the motional quadrature $X_{\rm S}$ of a mechanical oscillator, which is coupled with laser via a damped cavity mode. Efficient information retrieval is achieved by injecting the sensor output (green wavepacket) into a negative mass oscillator. As a result, the output field of the negative mass oscillator (blue wavepacket) contains only the signal and the shot noise, with the backaction noise canceled. (b) The signal manifests as a shift $\propto f$ of the mean quadrature $\langle P_{\rm E}(t)\rangle$ of the output field, which can be read out via homodyning~\cite{Moller:2017ta,PhysRevLett.105.123601,PhysRevLett.121.103602,PhysRevLett.121.031101}. Here $X_{\rm E}(t)=[b(t)+b^\dag(t)]/\sqrt{2}$ and $P_{\rm E}(t)=i[b^\dag(t)-b(t)]/\sqrt{2}$, the blurred disk denotes the laser shot noise. (c) Our general method, when applied to linear sensors, consists of a negative mass oscillator displaced by the prior information $f_0$, which results in an output field slightly shifted from the vacuum. Counting the output field achieves the same precision as (b). }
\label{fig:fig2} 
\end{figure}

The sensor further couples to the probe light (e.g., the waveguide mode) via its quadrature $X_{\rm S}$, as described by a jump operator $J_{\rm S}=\sqrt{\Gamma} X_{\rm S}$ with $\Gamma$ the coupling strength. For an optomechanical force sensor, as illustrated in Fig.~\ref{fig:fig2}, such a coupling is mediated by a driven damped cavity mode $(c, c^\dag)$, which interacts with the oscillator via the linearized optomechanical coupling $V= g (c^\dag + c) X_{\rm S}$. Assuming the cavity damping rate $\kappa\gg g$, adiabatic elimination of the cavity mode results in the aforementioned quadrature coupling $J_{\rm S}$ with $\Gamma=g^2/\kappa$. For an atomic gas magnetometer, such coupling is enabled by the Faraday rotation~\cite{Moller:2017ta}. 

Given $H_{\rm S}$ and $J_{\rm S}$, the decoder parameters can be determined via Eq.~\eqref{eq:H_J_general_solution} by solving for the sensor state $\rho_{\rm S}(t)$. This is particularly simple in the present linear oscillator case, as the sensor state is a Gaussian state. Straightforward calculation (see Appendix \ref{app: linear_sensor} for details)
provides 
\begin{equation}
\label{eq:negmassH}
H_{\rm D} = -\frac{1}{2}\omega (X_{\rm D}^2 + P_{\rm D}^2)-f X_{\rm D}
\end{equation}
and $J_{\rm D}=\sqrt{\Gamma}X_{\rm D}$. Equation \eqref{eq:negmassH} describes a negative frequency (or equivalently, negative mass) oscillator displaced by the prior information $f_0$. Experimentally, such an oscillator can be implemented effectively, e.g., with an ancillary atomic ensemble with properly adjusted detuning~\cite{Moller:2017ta}. By displacing the ancillary oscillator by the prior information of the unknown force $f_0$, counting the oscillator output reveals slight difference between the unknown force $f$ and its prior value $f_0$, cf. Fig.~\ref{fig:fig2}(c), allowing for achieving the QCRB in the neighborhood of $f_0$.

For linear sensors, information of the unknown force $f$ is contained only by the output \emph{signal}, i.e., by the mean quadrature of the emission field, not by its \emph{quantum noise}. This allows for achieving the QCRB with a simpler design, corresponding to setting $f=0$ in Eq.~\eqref{eq:negmassH} for the ancillary oscillator, and homodyne detection of its output, cf. Fig.~\ref{fig:fig2}(b), as in standard quantum noise cancellation schemes~\cite{PhysRevLett.102.020501,Moller:2017ta,PhysRevLett.105.123601,PhysRevLett.121.103602,PhysRevLett.121.031101}. For generic nonlinear sensors, however, the full output field, including the signal as well as the quantum noise, contains useful information---the QCRB can only be achieved by exploiting the full counting statistics of the decoder output, as we illustrate below.

\subsection{Nonlinear Quantum Sensors}
\label{sec:nonlinearSensor}
\begin{figure*}[t!]
\centering{} \includegraphics[width=1\textwidth]{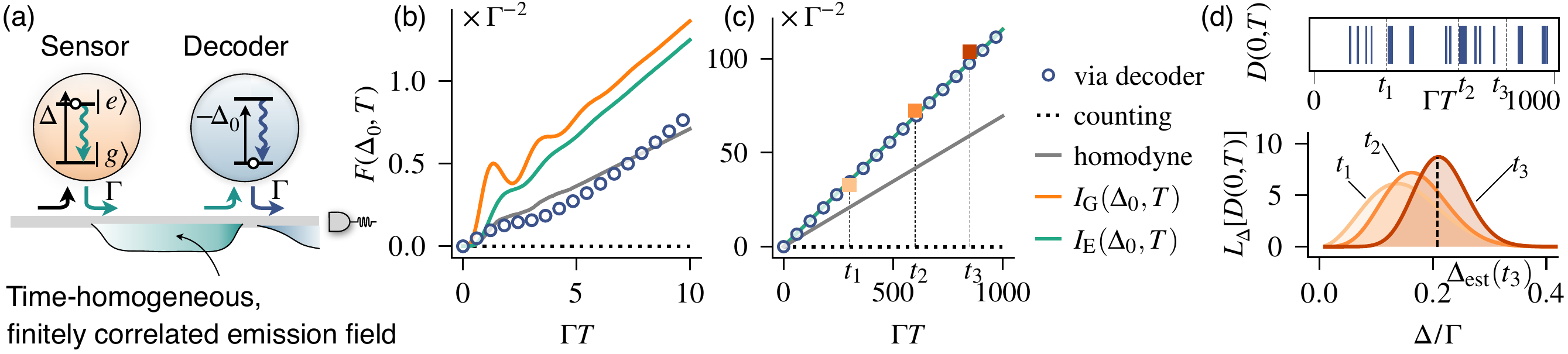} 
\caption{Efficient information retrieval for a nonlinear two-level sensor driven to the stationary state, of which the emission field is time-homogeneous and finitely correlated. As a demonstration we consider the sensing of the emitter detuning $\Delta$. (a) Efficient retrieval is enabled by choosing the decoder detuning opposite to the prior information of the sensor detuning $\Delta_0$, and other parameters identical. (b-c) The short-time (b) and long-time (c) behavior of the FI retrieved via the decoder, in comparison with the FI of direct photon counting, the FI of homodyne measurement (optimized over the homodyne angle), the global QFI $I_{\rm G}(\Delta_0,T)$, and the QFI of the emission field $I_{\rm E}(\Delta_0,T)$. Parameters: $\Omega=3\Gamma,\Delta_0=0$. For these parameters, the optimal homodyne angle is $\pi$. The sensor is initialized in $\ket{g}$. (d) Simulation of a single interrogation of our decoder-assisted sensing scheme. We consider that the sensor detuning slightly deviates from its prior value, $\Delta=\Delta_0+\delta$ with $\delta=0.2\Gamma$. Processing the continuous measurement record $D(0,T)$ (upper panel) provides the likelihood function ${L}_{\Delta}[D(0,T)]$, shown for $T\in\{t_1,t_2,t_3\}=\{300,600,850\}\times\Gamma^{-1}$ respectively. A maximum likelihood strategy provides us with an estimator $\Delta_{\rm est} (T)$, shown for $T=t_3$. The normalized inverse variance via averaging the estimators of $K = 5 \times10^3$ independent interrogations is shown as squares in (c).}
\label{fig:fig3} 
\end{figure*}
Intrinsically nonlinear quantum-optical devices, such as natural or synthetic few-level emitters interfaced with light~\cite{Northup:2014tt,RevModPhys.87.347,RevModPhys.87.1379,RevModPhys.89.021001}, are emergent platforms for the generation of large-scale multi-photon entangled states~\cite{PhysRevLett.95.110503,PhysRevLett.120.130501,PhysRevLett.128.010607,Besse:2020vq}, promising for sensing. We illustrate our retrieval scheme using two representative sensor designs in this context. First, a stationary nonlinear two-level emitter that emits temporally homogeneous field, of which the retrieved FI grows linearly with time. Second, a nonlinear three-level emitter subject to time-varying driving and features highly correlated emission field, of which the retrieved FI grows quadratically with time (i.e., obeying the Heisenberg scaling). These basic examples can be extended to more complex sensor designs, e.g.,  clusters of emitters.

\subsubsection{Stationary nonlinear emitters}
\label{sec:TLS}
Consider a driven two-level emitter coupled with a waveguide, cf. Fig.~\ref{fig:fig3}(a). Denoting the excited(ground) state as $\ket{e(g)}$, we write the Hamiltonian of the emitter in the rotating frame as
\begin{equation}
\label{eq:TLS_sensor_H}
H_{\rm S}=-\Delta \sigma_{ee}+\frac{1}{2}(\Omega \sigma_{eg}+{\rm h.c.}),
\end{equation}
where $\Delta$ and $\Omega$ (assumed real) are respectively the detuning and the Rabi frequency of the classical drive (e.g., laser), and we have used the notation $\sigma_{ij}=\ket{i}\bra{j}$. The emitter couples to the waveguide at a rate $\Gamma$ via the channel $\ket{e}\to\ket{g}$, as captured by the jump operator $J_{\rm S}=\sqrt{\Gamma_{\rm}}\sigma_{ge}$.

The LME ~\eqref{eq:ME} of the open sensor reduces to the familiar optical Bloch equation. Initialized in an arbitrary state $\ket{\psi_{\rm S}(0)}$, the emitter quickly relaxes, for $t\gtrsim \Gamma^{-1}$, to its (unique) stationary state $\rho_{\rm S}^{\rm st}=[s\sigma_{ee}+(s+2)\sigma_{gg}+\sqrt{2s}(e^{i\theta}\sigma_{eg}+e^{-i\theta}\sigma_{ge})]/2(1+s)$, where we have defined the saturation parameter $s=2\Omega^2/(4\Delta^2+\Gamma^2)$ and the angle $\theta={\rm arctan}(\Gamma/2\Delta)$. The state of the emission field is determined by $H_{\rm S}$ and $J_{\rm S}$ via Eqs.~\eqref{eq:KrausOP} and \eqref{eq:MPDOenv}, featuring a matrix-product structure of bond dimension $D=2$. For $t\gtrsim \Gamma^{-1}$, the boundary effect associated with $\ket{\psi_{\rm S}(0)}$ diminishes and the emission field is well approximated by a time-translational invariant MPDO.

For such a sensor model, the performance of conventional continuous measurement schemes are studied in~\cite{PhysRevLett.112.170401,PhysRevA.89.052110,PhysRevA.94.032103} and are shown to be generally suboptimal. A notable exception is the sensing of the Rabi frequency $\Omega$ of a resonantly driven ($\Delta=0$) two-level sensor, for which direct homodyne measurement is able to retrieve the QFI of the emission field~\cite{PhysRevA.94.032103}.

To efficiently retrieve the QFI for the sensing of any quantity in all parameter regimes, we rely on the decoder constructed according to our time-stationary recipe Eq.~\eqref{eq:diss_dimer}. Direct diagonalization of the stationary state provides us with $\rho_{\rm S}^{\rm st}=\sum_{k=\pm}p_k \ket{k}\bra{k}$, with $p_{\pm}=1/2\pm\sqrt{2s+1}/2(1+s)$ and $\ket{\pm}=[\sqrt{s}(\sqrt{1+2s}\mp 1)^{-1/2}\ket{g}\pm e^{i\theta}(\sqrt{1+2s}\mp 1)^{1/2}/\sqrt{2}\ket{e}]/(1+2s)^{1/4}$. Equation~\eqref{eq:diss_dimer} thus reduces to $H_{\rm D}=-W_0 H_{\rm S} W_0^\dag$ and $J_{\rm D}=-W_0 J_{\rm S} W_0^\dag$, with $W_0$ an arbitrary unitary. This provides us with an infinite number of possible choices of the decoder parameters $(H_{\rm D},J_{\rm D})$, as connected via the global gauge freedom $W_0$, all eligible to optimally retrieve the QFI of the emission field. In an actual experimental implementation one can take advantage of such a freedom to construct the easiest possible implementation of the decoder. For definiteness, we will make the choice $W_0=\sigma_{ee}-\sigma_{gg}$ and as a result
\begin{equation}
\label{eq:TLS_HD}
H_{\rm D}=\Delta \sigma_{ee}+\frac{1}{2}(\Omega\sigma_{eg}+{\rm h.c.}),
\end{equation}
and $J_{\rm D}=J_{\rm S}=\sqrt{\Gamma}\sigma_{ge}$, which requires an implementation of the decoder identical to the sensor, except for a detuning of opposite sign. {The cascaded sensor-decoder configuration described by Eqs.~\eqref{eq:TLS_sensor_H} and~\eqref{eq:TLS_HD} has been shown to establish exotic dark pure entangled states (cf.~Sec.~\ref{sec:sensor-decoder-entanglement})---the dissipative quantum spin dimers~\cite{Stannigel_2012,PhysRevLett.113.237203,PhysRevA.91.042116}---in the context of dissipative phase engineering.} Here we focus on its use for optimal information retrieval from the emission field.

We consider the sensing of the detuning of the emitter for illustration, $\theta=\Delta$. For efficient information retrieval, the decoder is detuned by the opposite of the prior information, $-\Delta_0$, as illustrated in Fig.~\ref{fig:fig3}(a). In Fig.~\ref{fig:fig3}(b) and Fig.~\ref{fig:fig3}(c), which correspond respectively to the short-time ($T\simeq$ a few $\Gamma^{-1}$) and the long-time ($T\gg \Gamma^{-1}$) regime of the sensor evolution, we plot and compare the relevant precision bounds, including the global QFI $I_{\rm G}(\Delta_0, T)$, the QFI of the emission field $I_{\rm E}(\Delta_0, T)$, the FI retrieved by direct counting and homodyning (optimized over all possible homodyne angles), and finally the FI retrieved via the decoder. The calculation of the FI for homodyne measurement is based on the numerical methods introduced in Refs.~\cite{PhysRevA.94.032103,Albarelli2018restoringheisenberg}.
As can be seen, after an initial transient oscillation, the precision bounds grow linearly with the interrogation time $T$---a typical behavior of time-stationary open sensors~\cite{PhysRevLett.106.090401,PhysRevLett.112.170401,2015JPhA...48J5301C}. Moreover, $I_{\rm G}(\Delta_0, T)$ and $I_{\rm E}(\Delta_0, T)$ differ by a small, time-independent constant that is hardly discernible for a long interrogation time, confirming the prediction of Sec.~\ref{sec:eQFI}. For the chosen simulation parameters, homodyne measurement is capable of retrieving a part of the QFI, whereas photon counting is completely ineffective~\footnote{It can be shown that $F(\Delta,T)$ is strictly zero for a resonantly driven two-level sensor. Indeed, photon counting can not resolve the sign of the small variation of $\Delta$ around $\Delta=0$. As a result, $\partial_{\Delta}P_{\Delta}[D(0,T)]|_{\Delta=0}=0$, resulting in a vanishing FI.}. 
In the short time limit, the FI retrieved by the decoder may not necessarily surpass the FI of conventional measurements, e.g., homodyne measurement, c.f., Fig.~\ref{fig:fig3}(b). This, however, does not contradict the long-time optimum of our retrieval scheme. The long-time optimum is evident from Fig.~\ref{fig:fig3}(c), where the FI retrieved by the decoder saturates $I_{\rm E}(\Delta_0, T)$ up to a small constant that is hardly discernible.

\begin{figure*}[t!]
\centering{} \includegraphics[width=1\textwidth]{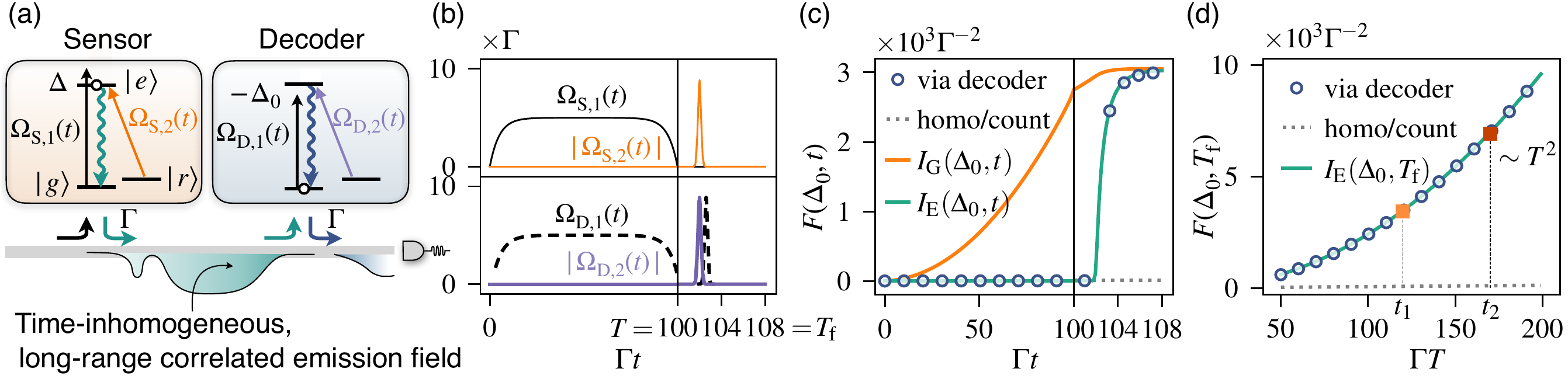} 
\caption{Efficient information retrieval for a nonlinear and nonergodic three-level sensor driven by a time-dependent protocol. (a) The sensor is driven by pulses $\Omega_{\rm S,1}(t)$ and $\Omega_{\rm S,2}(t)$ and emits correlated time-inhomogeneous multi-photon field featuring Heisenberg-limited QFI. To fully retrieve the QFI, the decoder is driven correspondingly by pulses $\Omega_{\rm D,1}(t)$ and $\Omega_{\rm D,2}(t)$. For other parameters of the decoder see text. (b) Pulse shapes for the emitter (upper panel) and the decoder (lower panel). (c) The FI retrieved via the decoder during the protocol, in comparison with the FI retrieved by direct counting and homodyning, the global QFI $I_{\rm G}(\Delta_0,t)$ and the QFI of the emission field $I_{E}(\Delta_0,t)$. We consider the sensing of the sensor detuning $\Delta$ as an example. Parameters: $\Delta_0=0,\Omega=5\Gamma$. (d) The FI retrieved via the decoder at the end of the protocol, in comparison with the FI retrieved by direct counting and homodyning, and the QFI of the emission field $I_{E}(\Delta_0,T_{\rm f})$ at the end of the protocol, for variable protocol duration $T$. Squares indicates the precision (the inverse of the variance) achieved by averaging the maximum likelihood estimation from $K = 5 \times10^3$ simulated sensing interrogations of period $T \in \{t_1, t_2\}=\{120,170\}\times \Gamma^{-1}$. Other parameters are the same as (c).}
\label{fig:fig4} 
\end{figure*}

As a demonstration of an actual sensing experiment, we consider that the sensor detuning slightly deviates from its prior value, $\Delta=\Delta_0+\delta$, and show in Fig.~\ref{fig:fig3}(d) the quantum trajectory simulation of a single repetition of the sensing of $\Delta$ using our decoder-assisted scheme. A simulated continuous signal $D(0,T)$, as provided experimentally by counting the output field of the decoder, is shown in the upper panel. Processing it via Eq.~\eqref{eq:SME} allows us to construct the likelihood function $L_\Delta[{D(0,T)}]:= \Gamma P_{\Delta}[D(0,T)]/\int d\Delta P_{\Delta}[D(0,T)]$, as shown in the lower panel for three different interrogation period $T=t_1,t_2,t_3$. A maximum likelihood strategy then provides us with an estimator $\Delta_{\rm est}(T)$ for the unknown detuning $\Delta$, which we show for $T=t_3$ as an example. Averaging the estimation of repeated interrogation reduces the sensing imprecision, as quantified by the variance ${\rm Var}[\Delta_{\rm est}(T)]=\langle[\Delta_{\rm est}(T)-\Delta]^2\rangle_{\rm st}$, with $\langle\cdot\rangle_{\rm st}$ denoting (classical) statistical average. We simulate $K=5\times10^3$ independent interrogations, for which the normalized inverse variance ${\rm Var}^{-1}[\Delta_{\rm est}(T)]/K$ is shown in Fig.~\ref{fig:fig3}(c) for three different interrogation period $T=t_1,t_2,t_3$. As can be seen, these match remarkably well with $F(\Delta_0,T)$, the theoretically asymptotic value in the $K\to \infty$ limit.

\subsubsection{Time-dependently driven nonlinear emitters}
\label{sec:3LS}
As an example of non-stationary and non-ergodic sensor designs, let us consider a driven-dissipative three-level emitter shown in Fig.~\ref{fig:fig4}(a), of which the two lower states $\ket{g},\ket{r}$ are driven to the excited state $\ket{e}$ by time-dependent fields $\Omega_{\rm S,1(2)}(t)$. We assume that the driving field for $\ket{g}\to\ket{e}$ is detuned from the transition frequency by $\Delta$, and $\ket{r}\to\ket{e}$ is driven on resonance. The sensor Hamiltonian can be written in the rotating frame as
\begin{equation}
H_{\rm S}(t)=\Delta \sigma_{gg}+\frac{1}{2}\left[\Omega_{\rm S,1}(t)\sigma_{eg}+\Omega_{\rm S,2}(t)\sigma_{er}+{\rm h.c.}\right].
\end{equation}
Besides the Hamiltonian evolution, we assume the sensor couples to the waveguide at a rate $\Gamma$ via the channel $\ket{e}\to\ket{g}$, as captured by the jump operator $J_{\rm S}=\sqrt{\Gamma_{\rm}}\sigma_{ge}$.

To generate correlated emission field featuring the Heisenberg scaling of the QFI, we initialize the sensor in a coherent superposition of the lower states, $\ket{\psi_{\rm S}(0)}=(\ket{g}-\ket{r})/\sqrt{2}$, and control the Rabi frequencies $\Omega_{\rm S,1(2)}(t)$ of the applied fields according to the envelop shown in the upper panel of Fig.~\ref{fig:fig4}(b): $\Omega_{\rm S,1}(t)=\Omega_{\rm S,1}\{{-\rm exp}(-t/\tau)-{\rm exp}[(t-T)/\tau]+{\rm exp}(-T/\tau)+1\}/[1+{\rm exp}(-T/\tau)-2{\rm exp}(-T/2\tau)]$, featuring gradual turning on at $t=0$ and off at $t=T$, and a constant strength in between [thin black line in Fig.~\ref{fig:fig4}(b)]; $\Omega_{\rm S,2}(t)$ is a Gaussian-shaped $\pi$ pulse of width $\sigma\ll \Gamma^{-1}$, applied at $t\gtrsim T+\Gamma^{-1}$ [thin orange line in Fig.~\ref{fig:fig4}(b)]. These time-dependent driving determine the state of the emission field as a temporally inhomogeneous MPDO via Eqs.~\eqref{eq:KrausOP} and \eqref{eq:MPDOenv}.

Such pulse envelops can be regarded as a continuous-time extension of the discrete scheme for the deterministic generation of photonic GHZ-type entanglement via quantum emitters~\cite{PhysRevLett.95.110503}, see also~\cite{PhysRevA.93.022103}. To develop a physical understanding, let us analyze the global state of the sensor and the environment $\ket{\Psi(t)}$. Initialized in $\ket{\Psi(0)}=(\ket{g}-\ket{r})\otimes \ket{\rm vac}/\sqrt{2}$, the global state evolves by sequential application of $A_{[n]}^{\sigma_n}$, constructed from $H_{\rm S}(t)$ and $J_{\rm S}$ via Eq.~\eqref{eq:KrausOP}. As $\ket{g}(\ket{r})$ is coupled(decoupled) with the waveguide during the time window $[0,T]$, and the sensor almost relaxes to $\ket{g}$ at time $t=T$, we have 
\begin{equation}
\label{eq:global_lambda}
\ket{\Psi(T)}\simeq(\ket{g}\otimes\ket{\rm bright}-\ket{r}\otimes \ket{\rm vac})/\sqrt{2},
\end{equation}
with $\ket{\rm bright}\simeq\sum_{\{\sigma_n\}} \bra{g}A_{[N]}^{\sigma_N}\dots A_{[1]}^{\sigma_1}\ket{g}\otimes\ket{\sigma_N\dots\sigma_2\sigma_1}$ an emission field MPS generated via the driven-dissipative dynamics in the channel $\ket{g}\leftrightarrow\ket{e}$ during $[0,T]$. Next, a short $\pi$ pulse $\Omega_{\rm S,2}(t)$ at phase $\pi/2$~\footnote{A $\pi$ pulse can be parameterized as $\Omega(t)\equiv e^{i\alpha}|\Omega(t)|$, with $\alpha$ being a time-independent global phase, and $|\Omega(t)|$ being its amplitude satisfying $\int|\Omega(t)|dt=\pi$} for $\ket{r}\to\ket{e}$ transforms the global state to $(\ket{g}\otimes\ket{\rm bright}-\ket{e}\otimes \ket{\rm vac})/\sqrt{2}$, and the emitter starts to relax $\ket{e}\to\ket{g}$ subsquently. At a final time $T_{\rm f}=T\,+$ a few $\Gamma^{-1}$,  the sensor and the emission field disentangles, $\ket{\Psi(T_{\rm f})}\simeq\ket{g}\otimes\ket{\Psi_{\rm E}(T_{\rm f})}$. The emission field state 
\begin{equation}
\label{eq:GHZ_env}
\ket{\Psi_{\rm E}(T_{\rm f})}\simeq(\ket{\rm bright}- \ket{\mathbf{1}_{[T,T_{\rm f}]}})/\sqrt{2} 
\end{equation}
features a GHZ-like superposition between $\ket{\rm bright}$, a multi-photon MPS generated in $[0,T]$, and $\ket{\mathbf{1}_{[T,T_{\rm f}]}}$, a single photon wavepacket emitted in $[T,T_{\rm f}]$. Importantly, unknown emitter parameters are encoded in $\ket{\rm bright}$ via the driven-dissipative dynamics of the sensor in $[0,T]$, with which the associated QFI can obey a Heisenberg scaling $\sim T^2$ with time. 

As a demonstration, we consider the sensing of the emitter detuning, $\theta=\Delta$, with a driving protocol involving realistic finite-width pulses, cf., Fig.~\ref{fig:fig4}(b). We rely on the general method introduced in Sec.~\ref{sec:eQFI} for the rigorous calculation of both the global QFI $I_{\rm G}(\Delta, t)$ and the QFI of the emission field $I_{\rm E}(\Delta, t)$; the results are shown in Fig.~\ref{fig:fig4}(c). As can be seen, in the time window $[0,T]$, the global QFI grows quadratically with time, whereas the QFI of the emission field grows linearly and stays small. The rapid growth of the global QFI is due to the strong entanglement between the sensor and the waveguide, cf. Eq.~\eqref{eq:global_lambda}; in contrast, the QFI of the reduced state of either the emission field or the sensor is small. This highlights the \emph{nonlocal} nature of the global QFI. Focusing on the reduced dynamics of the sensor, the different scaling behavior of the two QFIs reflects that the underlying sensor evolution in $[0,T]$ has multiple stationary states, cf. Sec. \ref{sec:QFIg}. These are easy to identify: a \emph{dark} stationary state $\rho_{\rm S}^{\rm st,1}=\ket{r}\bra{r}$ that is completely decoupled from the driven-dissipative channel $\ket{g}\leftrightarrow\ket{e}$; and a bright stationary state $\rho_{\rm S}^{\rm st,2}\simeq\ket{g}\bra{g}$ established by the driven-dissipative channel $\ket{g}\leftrightarrow\ket{e}$. After the time window $[0,T]$, the application of the short $\pi$ pulse $\Omega_{{\rm S},2}(t)$ and subsequent relaxation of the emitter in $[T,T_{\rm f}]$ disentangles the sensor and the waveguide. As a result, the global QFI is transferred into the emission field state Eq.~\eqref{eq:GHZ_env}, as evident from the sharp growth of the emission-field QFI $I_{\rm E}(\Delta_0,t)$ in Fig.~\ref{fig:fig4}(c). Seeing from the perspective of the reduced dynamics of the sensor, the short pulse $\Omega_{\rm S,2}(t)$ and subsequent relaxation pumps the population of $\rho_{\rm S}^{\rm st,1}$ into $\rho_{\rm S}^{\rm st,2}$, resulting in a single stationary state (thus recovering ergodicity) at the end of the sensing protocol.

To efficiently retrieve the Heisenberg-limited QFI, we construct the parameters of the decoder according to our general recipe Eq.~\eqref{eq:H_J_general_solution} by propagating the LME of the sensor. The resulting evolution of the decoder is parameterized by a time-dependent Hamiltonian
\begin{equation}
H_{\rm D}(t)=-\Delta \sigma_{gg}+\frac{1}{2}\left[\Omega_{\rm D,1}(t)\sigma_{eg}+\Omega_{\rm D,2}(t)\sigma_{gr}+{\rm h.c.}\right],
\end{equation}
and a time independent jump operator $J_{\rm D}=J_{\rm S}\equiv\sqrt{\Gamma}\sigma_{ge}$ during the time window spanning from a few $\Gamma^{-1}$ until $T$~\footnote{In the present case, to capture the single photon emitted by the sensor during $[T,T_{\rm f}]$, the decoder-waveguide coupling should be turned off gradually after the application of $\Omega_{\rm S,2}(t)$ and $\Omega_{\rm D,2}(t)$, i.e., $J_{\rm D}(t)\simeq J_{\rm S}e^{-\Gamma (t-T)/2}$ for $t\gg T$, similar to scenarios of quantum state transfer~\cite{PhysRevLett.78.3221}}. The temporal envelope of the driving pulses $\Omega_{\rm D,1(2)}(t)$ are shown in the lower panel of Fig.~\ref{fig:fig4}(b). The pulse $\Omega_{\rm D,1}(t)$ (thick dashed line) is almost identical to $\Omega_{\rm S,1}(t)$ and features gradual turning on and off; it further includes a short $\pi$ pulse (at phase $\pi/2$) in $[T,T_{\rm f}]$ immediately after $\Omega_{\rm S,2}(t)$. The short $\pi$ pulse $\Omega_{\rm D,2}(t)$ (at phase zero) is applied at the same time as $\Omega_{\rm S,2}(t)$. In Fig.~\ref{fig:fig4}(c), the FI retrieved via the decoder is compared with the QFI of the emission field, the remarkable match confirms the optimality of our retrieval scheme. In Fig.~\ref{fig:fig4}(d), we show the QFI of the emission field at the end of the protocol, $I_{\rm E}(\Delta_0,T_{\rm f})$, and the FI retrieved by our scheme, $F(\Delta_0,T_{\rm f})$, for variable duration $T$ (with $T_{\rm f}-T$ kept fixed), which demonstrate a remarkable Heisenberg scaling. We further show the precision (the inverse of the variance) achieved by averaging the estimators via the maximum likelihood strategy for $K = 5 \times10^3$ simulated sensing interrogations of various period $T=t_1,t_2$, which matches the retrieved FI well. Finally, we emphasize the crucial role of the ergodicity-recovering pulses of this driving protocol in retrieving the Heisenberg limited QFI.

In contrast, direct counting or homodyning of the emission field only retrieves a vanishing portion of the QFI, and does not show super-linear scaling, cf. Fig.~\ref{fig:fig4}(c-d). This can be understood by adopting the physical picture developed above: counting (or any other time-local measurement) completely destroys the coherence between the two components $\ket{\rm bright}$ and  $\ket{\mathbf{1}_{[T,T_{\rm f}]}}$ in the state of the emission field Eq.~\eqref{eq:GHZ_env}, thus losing the advantageous precision scaling.

\subsection{Many-body Quantum Sensors}
\label{sec:many-body-sensing}
\begin{figure}[t!]
\centering{} \includegraphics[width=0.48\textwidth]{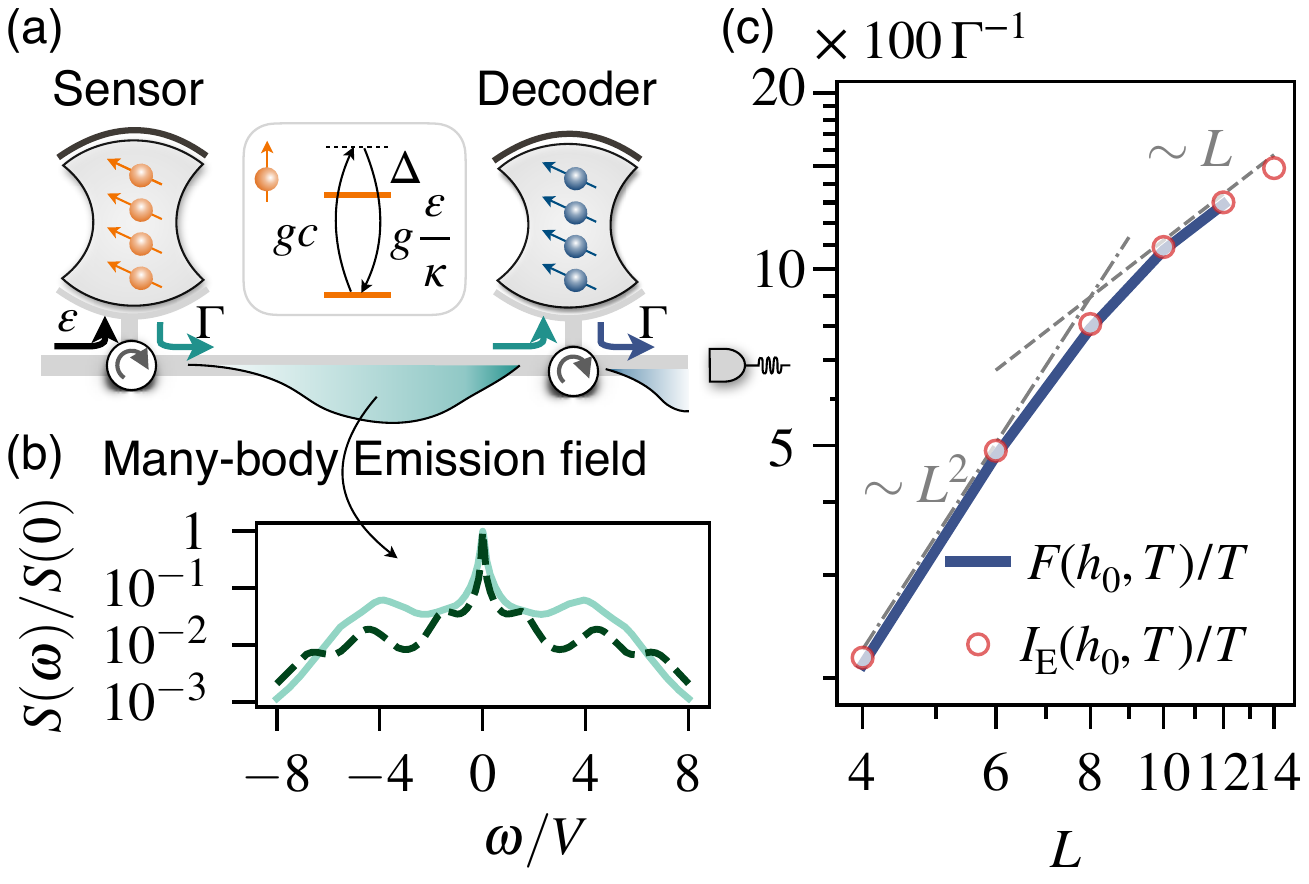} 
\caption{Efficient information retrieval for a driven-dissipative many-body sensor consisting of a transverse-field Ising spin chain. (a) The decoder can be realized as a replica spin chain with an opposite transverse field, cf. Eq.~\eqref{IsingHD}. The interface between the sensor (decoder) and the waveguide can be realized by coupling the spins to a driven damped cavity mode, a resonant contribution of such coupling is shown in the inset. (b) The autocorrelation spectrum of the sensor emission field for different transverse-field strength $h$: $h=0.5V$ (light solid line) and $h=V$ (dark dashed line). Other parameters: $L=8,\Gamma=0.1V,\alpha=3$. (c) The FI for the sensing of the transverse field $h$ retrieved via the decoder, $F(h_0, T)$, in comparison with the QFI of the emission field $I_{\rm E}(h_0, T)$, for different size $L$ of the many-body sensor. Parameters: $h_0=4V, \Gamma=V, \alpha=\infty$ (i.e., nearest neighbor interaction).}
\label{fig:fig5} 
\end{figure}

Finally, we demonstrate our information retrieval scheme with an open quantum many-body sensor consisting of a driven-dissipative transverse-field Ising spin chain, cf., Fig.~\ref{fig:fig5}(a). The Hamiltonian of the spin chain is
\begin{equation}
\label{IsingH}
H_{\rm S}= -\sum_{i\neq  j}^L \frac{V}{|i-j|^\alpha}\sigma^x_i\sigma^x_j-h\sum_{i=1}^L\sigma^z_i.
\end{equation}
We consider that the open spin chain couples to an optical fibre as its input-output channel, via the jump operator $J_{\rm S}=\sqrt{\Gamma}\sum_i \sigma^z_i$.

Such a sensor design can be realized naturally, e.g., with multiple ions trapped in a linear Paul trap, of which the ion number $L$ typically ranges from a few to a few tens~\cite{doi:10.1126/science.abg8102,doi:10.1126/science.abk2400}. The transverse-field Ising Hamiltonian Eq.~\eqref{IsingH} can be engineered via driving the internal state of the ions (i.e., the spins) with far-detuned global M{\o}lmer-S{\o}rensen beams, which provides adjustable exponent $\alpha$ for the spin-spin coupling, as routinely done in ion-based analog quantum simulation~\cite{doi:10.1126/science.abg8102,doi:10.1126/science.abk2400}. The ion-fibre interface can be mediated via an optical cavity, cf. Fig.~\ref{fig:fig5}(a), as realized experimentally~\cite{PRXQuantum.2.020331}. We consider driving the cavity mode with a coherent input with strength $\varepsilon$, and consider the regime $\Delta\gg\kappa\gg g,\varepsilon$, where $\Delta$ is the cavity-ion detuning, $\kappa$ is the cavity damping rate, and $g$ is the cavity-ion coupling strength. In this regime, the spin chain and the cavity mode interact via a linearized dispersive coupling $H_{\rm coup}=-\varepsilon g^2 (c+c^\dag) \sum_i \sigma^z_i/(\Delta{\kappa})$. Adiabatic elimination of the cavity mode results in the desired $J_{\rm S}$, with an effective decay rate $\Gamma=4\varepsilon^2g^4/(\Delta^2\kappa^3)$.

\begin{figure*}[t!]
\centering{} \includegraphics[width=1\textwidth]{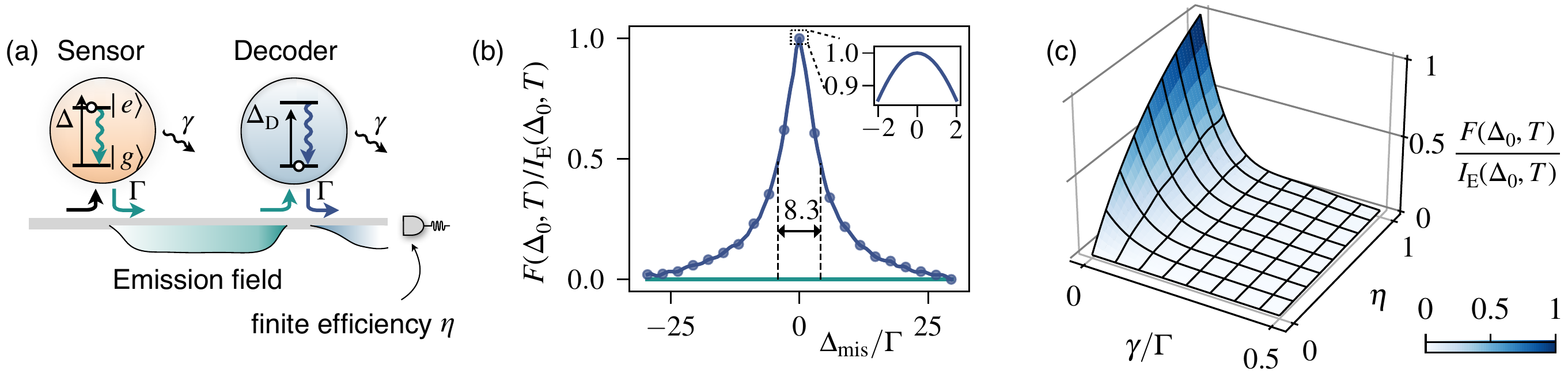} 
\caption{Robustness of the information retrieval scheme against experimental imperfections demonstrated via the nonlinear two-level sensor example. (a) Experimental imperfections include the finite accuracy in the control of the decoder parameters, light transmission loss (at a rate $\gamma$), finite detector efficiency $\eta<1$, and dephasing of the sensor and decoder (at a rate $\gamma$). (b) The FI retrieved for finite accuracy in the control of the decoder parameters, examplified by a mismatch $\Delta_{\rm mis}$ between the decoder detuning $\Delta_{\rm D}$ and its ideal value $-\Delta_0$, $\Delta_{\rm mis} = \Delta_{\rm D} +\Delta_0$. The horizontal line shows the FI retrieved by direct photon counting (without the decoder). Parameters: $\Omega=\Gamma,\Delta_0=\gamma=0,\eta=1$. (c) The FI $F(\Delta_0,T)$ retrieved for finite $\eta<1$ and $\gamma>0$, assuming perfect control of the decoder, $\Delta_{\rm mis}=0$. Parameters for the simulation: $\Omega=\Delta_0=\Gamma$.}
\label{fig:fig6} 
\end{figure*}

We consider the sensor operating in the stationary regime, and construct the decoder parameters via Eq.~\eqref{eq:diss_dimer}. The unique stationary state of the sensor in our case is an infinite temperature state $\rho_{\rm S}^{\rm st}=\mathbb{I}_{D}/D$ with $D=2^L$, which allows for the reduction of Eq.~\eqref{eq:diss_dimer} to a remarkably simple relation, $H_{\rm D}=-W_0 H_{\rm S} W_0^\dag$ and $J_{\rm D}=-W_0 J_{\rm S} W_0^\dag$, with $W_0$ an arbitrary unitary. We emphasize that such a relation holds for a broad class of (finite-size) many-body sensors: that are described by a frustration-free Hamiltonian $H_{\rm S}$ and a Hermitian jump operator $J_{\rm S}$ which mutually do not commute, $[H_{\rm S},J_{\rm S}]\neq0$, such that its unique stationary state is an infinite temperature state. Note that while such a stationary state is insensitive to external unknown parameters, the emission field of the sensor is highly sensitive. Indeed, the emission field inherits the complex temporal correlations from the quantum many-body dynamics as governed by $H_{\rm S}$. We illustrate this in Fig.~\ref{fig:fig5}(b) in terms of the emission spectrum, $S(\omega)=\int dt e^{-i\omega t}\langle J_{\rm S}(t)J_{\rm S}(0)\rangle_{\rm st}$ [where $\langle\cdot\rangle\equiv{\rm tr}(\,\cdot\, \rho_{\rm S}^{\rm st})$], which is directly proportional to the dynamic structure factor $\int d t e^{-i\omega t}\sum_{ij}\langle\sigma_{i}^z(t)\sigma_{j}^z(0)\rangle_{\rm st}$ of our driven-dissipative spin model.

Choosing $W_0={\rm exp}(i\pi \sum_i \sigma_i^y/2)$ provides us with a realization of the decoder parameters
\begin{equation}
\label{IsingHD}
H_{\rm D}= -\sum_{i\neq j}^L\frac{V}{|i-j|^\alpha}\sigma^x_i\sigma^x_j+h\sum_{i=1}^L\sigma^z_i.
\end{equation}
and $J_{\rm D}=\sqrt{\Gamma}\sum_i \sigma^z_i$, which can be easily implemented with a replica setup of the sensor. In Fig.~\ref{fig:fig5}(c) we show the FI retrieved via the decoder for the sensing of the transverse field $h$ for different size $L$ of the many-body sensor, which matches the QFI of the emission field $I_{\rm E}(h_0, T)$ remarkably well. With increasing size $L$, the retrieved FI manifests a gradual crossover from a super-linear scaling $\sim L^2$ to a linear scaling $\sim L$. This can be qualitatively explained by noticing that the spin system under consideration possesses a finite correlation length $\xi$, therefore $F(h,T)\lesssim I_{\rm E}(h,T)\sim L^2$ for $L\lesssim \xi$, and $F(h,T)\lesssim I_{\rm E}(h,T)\sim \xi L$ for $L\gtrsim \xi$.

Beyond finite-range correlated lattice spin models, our information retrieval scheme can be applied to other open many-body sensor designs, e.g., sensors that operate in the vicinity of dissipative phase transitions and possess a divergent correlation length~\cite{PRXQuantum.3.010354}, which may enable potential precision enhancement based on dynamic critical phenomena.

\section{Robustness of the retrieval scheme against experimental imperfections}
\label{sec:imperfect}
Realistic implementations of our retrieval scheme is necessarily accompanied by imperfections. These include (i) inaccurate control of the decoder parameters, (ii) photon transmission loss, (iii) imperfect photon detection, e.g., finite detector efficiency and dark counts, and (iv) other decoherence channels such as sensor (decoder) dephasing. Nevertheless, the temporally quasi-local, matrix-product structure of our measurement scheme offers it remarkable resilience against the experimental imperfections, competitive with temporally local measurement. Below we demonstrate this at the hand of the nonlinear two-level sensor design introduced in Sec.~\ref{sec:nonlinearSensor}.

First, we examine the robustness of our scheme against imperfection (i), i.e., finite accuracy in the control of the decoder parameters. As a demonstration, we consider a mismatch between the decoder detuning $\Delta_{\rm D}$ and its ideal value 
$-\Delta_0$, $\Delta_{\rm mis}=\Delta_{\rm D}+\Delta_0$, and show in Fig.~\ref{fig:fig6}(b) the retrieved FI by the imperfect decoder. Remarkably, the imperfect decoder retrieves a significant portion of the QFI of the emission field over a broad range of $\Delta_{\rm mis}$, with a full width at half maximum (FWHM) $\simeq 8.3\Gamma$. We note that for sufficiently large $\Delta_{\rm mis}$, the decoder is far detuned from the sensor and cease to scatter the incoming radiation field. As a result, decoder-assisted measurement of the emission field approaches direct counting, providing the same FI [horizontal line in Fig.~\ref{fig:fig6}(b)]. The retrieved FI decreases slowly and remains close to the QFI for small $\Delta_{\rm mis}$, cf., inset of Fig.~\ref{fig:fig6}(b). These demonstrate remarkable robustness of our information retrieval scheme against uncertainties in the experimental control of the decoder.

Second, we demonstrate the robustness of our scheme against imperfection (ii-iv). Photon transmission loss, due to either waveguide attenuation or the imperfect coupling between the sensor (decoder) and the waveguide, can be modelled as decoherence via the introduction of additional (unmonitored) environments~\cite{PhysRevX.7.011035}. This result in additional decoherence terms ${\cal D}[J^{\rm S(D)}_{\rm dec}]\tilde{\varrho_{\rm}}_{c}dt$ to Eq.~\eqref{eq:SME}, with $J_{\rm dec}^{\rm S(D)}$ the associated jump operator of the sensor(decoder). To be specific, we consider $J_{\rm dec}^{\rm S(D)}=\sqrt{\gamma}\sigma_{ge}$, i.e., we assume the effective cooperativity of the sensor(decoder)-waveguide coupling to be ${\cal{C}}=\Gamma/\gamma$. Similary, the finite detector efficiency $\eta<1$ can be accounted by splitting the decoder output as detectable and undetectable parts, as modelled by an additional term $(1-\eta)J\tilde{\rho}_c J^\dag dt[1-d{\cal N}(t)]$ to Eq.~\eqref{eq:SME}. Finally, we include the dephasing of the sensor and decoder via the Lindblad operators $\gamma_{\rm dep}{\cal D}[\sigma_{z}^{\rm S(D)}]$, and we choose $\gamma_{\rm dep}=\gamma$ for illustration. 

We demonstrate in Fig.~\ref{fig:fig6}(c) the performance of our scheme under imperfections (ii-iv), by comparing the retrieved FI to $I_{\rm E}(\Delta_0,T)$, i.e., the retrieved FI in the absence of imperfections. As can be seen, finite detector efficiency $\eta<1$ and nonzero $\gamma/\Gamma$ in general reduces the retrieved FI. Nevertheless, at modest strength of the imperfections, $\gamma/\Gamma\leq 0.2$ and $\eta>0.3$, our scheme is able to extract a significant portion of $I_{\rm E}(\Delta_0,T)$, which demonstrates its experimental robustness. We note that these numbers are well within the reach of state-of-the-art platforms for quantum matter-light interface~\cite{doi:10.1063/1.4838696,PhysRevLett.110.243602,Hacker:2016ti,Landig:2018ta,Mirhosseini:2019uf,doi:10.1073/pnas.1603788113,doi:10.1126/science.aaj2118,Lodahl:2017vb,doi:10.1126/science.abi9917}.

\section{Conclusion and Outlook}
\label{sec:concusion}
We have established a general method that retrieves the full Quantum Fisher information (QFI) of the non-classical, temporally correlated fields emitted by generic quantum sensors under continuous measurement. As a result, our method allows for achieving the Quantum Cr\'amer Rao Bound (QCRB) for generic open quantum sensor designs. The key element in our method is a quantum decoder which transforms temporally correlated multi-photon states that are described by matrix product states (MPS) to simple product states at its output. By injecting the emission field of the open quantum sensor into the decoder and then performing conventional, time-local measurement on its output, our method effectively implements measurement of the sensor emission field in temporally non-local MPS bases. Appropriate design of the decoder, for which we construct a universal recipe via Eq.~\eqref{eq:H_J_general_solution} for generic open quantum sensors, renders the effective measurement capable of achieving the QCRB. We have illustrated the effectiveness of our method at the example of paradigmatic open sensor designs ranging from linear force sensors, via nonlinear emitters, to driven-dissipative many-body sensors, and we have examined the resilience of our method against experimental sources of noise and imperfections. With this verification of the feasibility, effectiveness and broad range of applicability, our method paves the way for improving a wide range of sensor platforms towards achieving their fundamental sensitivity limit allowed by quantum mechanics. Along with the information retrieval method, we have introduced an MPS framework for the description of continuous-measurement based sensing technology, which in particular results in an effective analytical expression for the evaluation of the QFI of the emission field of generic open sensors. This provides us with a refined, tighter QCRB for open quantum sensors as compared to existing bounds~\cite{PhysRevLett.106.090401,PhysRevLett.112.170401,2015JPhA...48J5301C}. The information retrieval method that we have introduced here can be viewed as a universal quantum backaction evasion strategy for generic (linear and nonlinear) driven-dissipative sensors. The establishment of such a strategy is timely and essential in view of the rapid progress of a broad spectrum of sensor technologies towards the realm where their precision is principally limited by the excess quantum noise of the sensor setups---our scheme provides a general strategy to optimally evade such noise.

Our method is particularly important for the emergent nonlinear and complex sensor platforms that have been developing rapidly in recent years in many areas of experimental quantum optics. One highlight of these is the synthetic quantum many-body systems integrated as sensors~\cite{Goban:2018aa,Marciniak:2022aa,Ding:2022aa} which, combined with our general method as demonstrated in Sec.~\ref{sec:examples}, may open up new vistas of driven-dissipative sensor devices that operate at their ultimate precision limit harnessing many-body correlations. Promising candidates ranges from ensembles of nitrogen-vacancy centers embedded in microwave resonators~\cite{Jin:2015vn} via driven-dissipative atomic gas~\cite{RevModPhys.85.553,Dreon:2022wd,doi:10.1126/science.abo3382} to many-body circuit QED~\cite{Ma:2019um} setups. Towards this goal, an attractive outlook for future research is the reduction of the complexity of the decoder design and maintaining at the same time the high information retrieval efficiency based on, e.g., low rank matrix-product-operator (MPO) approximations~\cite{Cramer:2010aa,PhysRevLett.111.020401,Lanyon:2017aa,Chabuda:2020tv} of the emission field of complex sensor designs.

A few other prospects are worth future exploration as well. First, our method can be readily extended beyond the determination of small variations around the prior value, to the estimation of parameters with a finite dynamical range. In the first place, this amounts to adaptive schemes via iterative adjustment of the decoder based on estimation from prior measurement. Optimization of the associated resource (e.g., time) consumption, e.g., by exploiting the machinery of Bayesian inference, is an interesting open question. Moreover, the extension of our scheme to optimal waveform detection~\cite{PhysRevA.86.042115,PhysRevLett.106.090401} is especially appealing in view of the natural compatibility between our scheme and time-dependent sensor designs. Finally, beyond the realm of quantum sensing, the quasi-local, finitely-correlated continuous measurement proposed in the present work may prove to be a powerful tool for other quantum technological applications including, e.g., the efficient certification and tomography of large-scale multi-photon entangled states for quantum simulation~\cite{PhysRevLett.110.090501,PhysRevX.5.041044,PhysRevLett.120.130501} and computation~\cite{doi:10.1073/pnas.1711003114}.

\section*{Acknowledgments}
D.~Y. acknowledges Ji-Yao Chen, Koenraad Audenaert and Theodoros Ilias for helpful discussions. We thank Benjamin D'Anjou for proofreading, and Madalin Guta and Peter Zoller for useful comments on the manuscript. This work was supported by the ERC Synergy grant HyperQ (Grant No.~856432) and the EU projects QuMicro (Grant No.~101046911). We acknowledge support by the state of Baden-Württemberg through bwHPC and the German Research Foundation (DFG) through grant no INST 40/575-1 FUGG (JUSTUS 2 cluster). Part of the numerical simulations were performed using the QuTiP library~\cite{JOHANSSON20131234} and the QuSpin package~\cite{10.21468/SciPostPhys.2.1.003}.

\emph{Note added:} While completing the manuscript, we became aware of a related work~\cite{Godley2023adaptivemeasurement} on achieving the quantum Cram\'er-Rao bound for discrete, homogeneous Markov chains via adaptive measurements. 

\appendix
\section{Quantum Fisher Information of the Emission Field---Proofs and Extensions}
In this appendix, we rigorously prove the formula for the efficient calculation of the QFI of the emission field presented in Sec.~\ref{QFIe}, and make a few extensions including generalization to the case of multiple environments.
\subsection{Proof of the Formula for the Quantum Fidelity of the Emission Field}
\label{eq: QFIe}
Here we prove the fidelity formula Eq.~\eqref{eq: envQFI_fomula}, which is the key for the calculation of the QFI of the emission field (environment). Noticing that the global state Eq.~\eqref{eq:MPSglobal} is an entangled state of the sensor and the environment, let us formally write down its Schmidt decomposition as 
\begin{equation}
\label{eq:schmidtD}
\ket{\Psi(\theta,T)}=\sum_{k=1}^D s_k(\theta,T) \ket{k(\theta,T)}_{\rm S}\otimes\ket{k(\theta,T)}_{\rm E}.
\end{equation} 
Here, $\ket{k(\theta,T)}_{\rm S(E)}$ are an orthonormal basis of the sensor (environment) (the Schmidt basis), and $s_k(\theta,T)$ are non-negative real numbers satisfying $\sum_{k=1}^D s_k^2(\theta,T)=1$ (the Schmidt coefficients), all dependent on the unknown parameter $\theta$. The reduced state of the sensor and the environment, as defined by Eqs.~\eqref{eq:ME} and \eqref{eq:MPDOenv} respectively, can therefore be expressed as $\rho_{\rm S(E)}(\theta,T)=\sum_{k=1}^D s_k^2(\theta,T)\ket{k(\theta,T)}_{\rm S(E)}\bra{k(\theta,T)}$, allowing us to represent $\sqrt{\rho_{\rm S(E)}}$ conveniently. As a result, the quantum fidelity of the environment state, cf. Eq.~\eqref{eq:quantum_fidelity_env}, can be expressed as
\begin{align}
\label{eq:F_E_app}
{\cal F}_{\rm E}(\theta_1,\theta_2)&={\rm tr_E}\bigg(\sqrt{\sum_{k,k''} f_{kk''}(\theta_1,\theta_2) \ket{k(\theta_1,T)}_{\rm E}\bra{k''(\theta_1,T)}}\bigg)\nonumber\\
&={\rm tr}\left(\sqrt{\boldsymbol{f}}\right),
\end{align}
in which we have defined the kernel 
\begin{align}
f_{kk''}(\theta_1,\theta_2)=&s_k(\theta_1,T)s_{k''}(\theta_1,T)\sum_{k'}\bigg(\!s_{k'}^2(\theta_2, T)\nonumber\\
&{}_{\rm E}\langle k(\theta_1,T)|k'(\theta_2,T)\rangle_{\rm E}\langle k'(\theta_2,T)|k''(\theta_1,T)\rangle_{\rm E}\!\bigg)\nonumber
\end{align} 
and the $D\times D$ matrix $\boldsymbol{f}$ via $\boldsymbol{f}_{kk''}=f_{kk''}(\theta_1,\theta_2)$.

Next, we relate Eq.~\eqref{eq:F_E_app} to the generalized density operator of the sensor $\mu_{\theta_1,\theta_2}(T)$, cf. Eq.~\eqref{eq:generalizedDM}. Adopting the Schmidt decomposition Eq.~\eqref{eq:schmidtD}, the generalized density operator can be expressed as
\begin{align}
\mu_{\theta_1,\theta_2}(T)
=&\sum_{kk'}s_{k}(\theta_1,T)s_{k'}(\theta_2,T)\,{}_{\rm E}\langle k'(\theta_2,T)|k(\theta_1,T)\rangle_{\rm E}\nonumber\\
&\times \ket{k(\theta_1,T)}_{\rm S}\bra{k'(\theta_2,T)}.
\end{align}
Straightforward calculation gives $\mu_{\theta_1,\theta_2}(T)\mu_{\theta_1,\theta_2}^\dag(T)=\sum_{k,k''}f_{k,k''}(\theta_1,\theta_2)\ket{k''(\theta_1,T)}_{\rm S}\bra{k(\theta_1,T)}$, in which $f_{k,k''}(\theta_1,\theta_2)$ is defined in the line under Eq.~\eqref{eq:F_E_app}. As such, its matrix representation is given by $\boldsymbol{f}^{\rm T}$ in the basis $\ket{k(\theta_1)}_{\rm S}$, with the superscript ${\rm T}$ denoting the matrix transpose. As trace is invariant under matrix transpose, we conclude 
 \begin{equation}
 {\cal F}_{\rm E}(\theta_1,\theta_2)={\rm tr}\left(\sqrt{\boldsymbol{f}^{\rm T}}\right)={\rm tr}\left[\sqrt{\mu_{\theta_1,\theta_2}(T)\mu_{\theta_1,\theta_2}^\dag(T)}\right],
 \end{equation}
which recovers Eq.~\eqref{eq: envQFI_fomula}.

\subsection{Equivalent Formulas for the Quantum Fisher Information}
Here we provide a few variants of Eqs.~\eqref{eq:QFIe} and \eqref{eq:QFIg} that relate the QFI to the quantum fidelity, which may be  convenient to use in certain cases. The discussion here is applicable to both $I_{\rm E}(\theta,T)$ and $I_{\rm G}(\theta,T)$. To simplify the notation, we will drop the subscripts and use $I(\theta, T)$ to represent the QFI and ${\cal F}(\theta_1,\theta_2)$ the corresponding fidelity.

First, by differentiating the identity ${\cal F}(\theta,\theta)=1$ and using the symmetry ${\cal F}(\theta_1,\theta_2)={\cal F}(\theta_2,\theta_1)$, it is easy to show $\partial_{\theta_1}\partial_{\theta_2} {\cal F}|_{\theta_1=\theta_2=\theta}=-\partial_{\theta_1}^2{\cal F}|_{\theta_1=\theta_2=\theta}$. This allows us to rewrite Eqs.~\eqref{eq:QFIe} and \eqref{eq:QFIg} in a symmetric form
\begin{equation}
I(\theta,T)=4\partial_{\theta_1}\partial_{\theta_2} {\cal F}(\theta_1,\theta_2)|_{\theta_1=\theta_2=\theta}.
\end{equation}

Second, we can make use of the fact that if an analytic function $f(\delta)$ satisfies $0\leq f(\delta)\leq f(0)=1$, $\forall \delta\in \mathbb{R}$, then $\partial_{\delta}^2{\rm log}f|_{\delta=0}=\partial_{\delta}^2f|_{\delta=0}$, as can be verified by straightforward calculation. Taking $f({\delta})={\cal F}(\theta,\theta+\delta)$, we can rewrite Eqs.~\eqref{eq:QFIe} and \eqref{eq:QFIg} as
\begin{equation}
\label{eq:QFI_alt_app}
I(\theta,T)=-4\partial_\delta^2 {\rm log}{\cal F}(\theta,\theta+\delta)|_{\delta=0}.
\end{equation}
Numerical calculation of the QFI requires the evaluation of ${\cal F}(\theta,\theta+\delta)$, which typically decreases with time exponentially for $\delta\neq 0$. In this regard, Eq.~\eqref{eq:QFI_alt_app} is numerically more robust than Eqs.~\eqref{eq:QFIe} and \eqref{eq:QFIg}, especially when $T$ is large.

\subsection{Extensions to Multiple Environments}
\label{app:A3}
In Sec.~\ref{QFIe}, we have discussed the formula for the QFI of the emission field, Eq.~\eqref{eq: envQFI_fomula}, under the assumption that the open quantum sensor is coupled with a single input-output channel. Here we point out that if the sensor is coupled with multiple environments, the same formula can be used to calculate the \emph{total} QFI of the environments, provided that the update law of the generalized density matrix $\mu_{\theta_1,\theta_2}(T)$ is appropriately adjusted. Specifically, consider that the sensor is coupled with environments $m=1,2,\dots M$ via the jump operator $J_{\rm S}^m$ and is interrogated by the Hamiltonian $H_{\rm S}(\theta,t)$ (which includes the possible Lamb shifts due to the sensor-environment coupling), the corresponding generalized density matrix evolves according to
\begin{align}
\label{eq:generalizedME2}
\frac{d\mu}{dt}=&-i\left[H_{\rm S}(\theta_1,t)\mu-\mu H_{\rm S}^\dag(\theta_2,t)\right]\nonumber\\
&+\frac{1}{2}\sum_{m=1}^M \bigg\{2J_{\rm S}^{m}(\theta_1,t)\mu J_{\rm S}^{m\dag}(\theta_2,t)\nonumber\\
&-\left[J_{\rm S}^{m\dag}(\theta_1,t) J_{\rm S}^m(\theta_1,t) \mu + \mu J_{\rm S}^{m\dag}(\theta_2,t) J_{\rm S}^m(\theta_2,t)\right]\bigg\}.
\end{align}
This, together with Eq.~\eqref{eq: envQFI_fomula}, provides us with the \emph{total} fidelity of the environments, that is, in its definition Eq.~\eqref{eq:quantum_fidelity_env} $\rho_{\rm E}(\theta,T)$ represents the joint state of all the environments. The proof of this is similar to that presented in Appendix \ref{eq: QFIe} by interpreting Eq.~\eqref{eq:schmidtD} as the Schmidt decomposition between the system (sensor) and all the environments it couples with.

Similarly, in the presence of multiple environments, the combination of Eq.~\eqref{eq:QFIg}, \eqref{eq:FG} and Eq.~\eqref{eq:generalizedME2} provides us with the global QFI of the sensor and all the environments it couples with, as established in Ref.~\cite{PhysRevLett.112.170401}.

\section{Efficient Information Retrieval---Proofs and Extensions}
In the main text, we have constructed the optimal effective measurement via the decoder based on physical argument. In this appendix, we rigorously justify such a construction. Specifically, we will prove that {\bf(I)} the `optimal' effective measurement of the sensor emission field, as introduced in Sec.~\ref{sec:optimal_meas} and defined by Eqs.~\eqref{eq:optimal_cond} and \eqref{eq:vac_projector}, is indeed optimal---it allows, in the neighborhood of the prior information, for the retrieval of the full QFI of the sensor emission field up to a (finite) constant; and {\bf (II)} such an effective measurement can be realized by designing the decoder evolution according to the central theorem of Sec.~\ref{sec:decoder_recipe}. Besides the two proofs, we will also discuss the sensor-decoder entanglement, the efficiency of the maximum likelihood estimation (MLE), and possible extension of our scheme to the decoding of spatially correlated many-body states described by discrete MPSs. 

\subsection{The Central Theorem for the Decoder Construction Realizes the Desired Effective Measurement}
\label{app: proof_left_normalization}
Let us first prove {\bf (II)}, i.e., the central theorem realizes Eqs.~\eqref{eq:optimal_cond} and \eqref{eq:vac_projector}. This is equivalent to prove that $P_{\theta}[\{\sigma_n^*\}]\equiv\langle\Psi_{\rm tot}(\theta,T)|\Pi_{\{\sigma^*_n\}}|\Psi_{\rm tot}(\theta,T)\rangle=1$ for $\theta=\theta_0$, where $\Pi_{\{\sigma_n^*\}}=\otimes_{n=1}^{N} |0\rangle_n\langle0|$ as defined in Eq.~\eqref{eq:vac_projector}, and $|\Psi_{\rm tot}(\theta,T)\rangle$ is the ($\theta$-dependent) joint state of the sensor, the environment and the decoder at time $T$. We can express $\ket{\Psi_{\rm tot}(\theta,T)}=U_{\rm tot}(\theta,T)\ket{\Psi_{\rm tot}(0)}$, with $U_{\rm tot}(\theta,T)={\cal T}U_{\rm DE}(T)U_{\rm SE}(\theta,T)$ the joint evolution operator of the sensor, the environment and the decoder, and $\ket{\Psi_{\rm tot}(0)}=\ket{\psi_{\rm S}(0)}\otimes\ket{\rm vac}\otimes\ket{\psi_{\rm D}(0)}$ their joint initial state. We note that $\theta$ enters the sensor-environment evolution $U_{\rm SE}(\theta,T)$, but not the decoder-environment evolution $U_{\rm DE}(T)$.

Let us further define an (unnormalized) state of the sensor and the decoder by projecting the total state $\ket{\Psi_{\rm tot}(\theta, T)}$ onto the environmental vacuum state,
\begin{equation}
\label{eq:psi_SD}
|\tilde{\psi}_{\rm SD}(\theta, T)\rangle :=\left(\otimes_{n=1}^N\langle0_n|\right)|\Psi_{\rm tot} (\theta,T)\rangle.
\end{equation}
It allows us to express $P_{\theta}[\{\sigma_n^*\}]=\langle \tilde{\psi}_{\rm SD}(\theta,T)|\tilde{\psi}_{\rm SD}(\theta,T)\rangle$. By exploiting the expression of $U_{\rm SE}(\theta, T) $ and $U_{\rm DE}(T) $, cf. Eq.~\eqref{eq: evolution_int}, we can express
\begin{align}
\label{eq:psi_SD_evolve}
|\tilde{\psi}_{\rm SD}(\theta,T)\rangle 
=&\sum _{\left\{\sigma _n\right\}} \left(\bar{B}_{[N]}^{\sigma _N}\dots\bar{B}_{[n]}^{\sigma _n}\dots\bar{B}_{[1]}^{\sigma _1}|\psi _{\rm D}(0)\rangle
\right)\nonumber\\
&\otimes \left(A_{[N]}^{\sigma _N}(\theta)\dots A_{[n]}^{\sigma _n}(\theta)\dots A_{[1]}^{\sigma _1}(\theta)|\psi _{\rm S}(0)\rangle \right),
\end{align}
in which the \(D\times D\) matrices read $A_{[n]}^{\sigma_n}(\theta)\equiv\langle \sigma_n | U_{\rm SE}^{[n]}(\Delta t)|0_n\rangle$ and similarly $\bar{B}_{[n]}^{\sigma_n}\equiv\langle 0_n | U_{\rm DE}^{[n]}(\Delta t)|\sigma_n\rangle$, $\sigma _n=0,1$. The tensors $A_{[n]}(\theta)$ are related to the ($\theta$-dependent) sensor Hamiltonian and jump operator via Eq.~\eqref{eq:KrausOP}, and similarly $\bar{B}_{[n]}$ are related to the decoder Hamiltonian and jump operator via Eq.~\eqref{eq:Kraus_anti}. Exploiting the Choi-Jamiolkowski isomorphism \(Y\otimes X \sum _{i j} c_{i j}|j\rangle \otimes |i\rangle \leftrightarrow X \sum _{i j} c_{i j}|i\rangle\langle j| Y^{\rm T}\), we further have $P_{\theta}[\{\sigma_n^*\}]=\text{tr}\left[\Lambda(\theta,T) \Lambda ^{\dagger }(\theta,T)\right]$ with 
\begin{align}
\Lambda(\theta,T) =&\sum _{\left\{\sigma _n\right\}} \left(A_{[N]}^{\sigma _N}(\theta)\dots A_{[n]}^{\sigma _n}(\theta)\dots A_{[1]}^{\sigma _1}(\theta)\right)|\psi _{\rm S}(0)\rangle\nonumber\\
&\times\langle \psi _{\rm D}(0)|\left(B_{[1]}^{\sigma _1\dag}\dots B_{[n]}^{\sigma _n\dag}\dots B_{[N]}^{\sigma _N\dag}\right).
\label{eq:choi_map_matrix}
\end{align}

Now we are ready to show that the choice of the initial state and the evolution of the decoder according to our central theorem guarantees that $P_{\theta}[\{\sigma_n^*\}]=1$ for $\theta=\theta_0$. 
With such a choice, in Eq.~\eqref{eq:choi_map_matrix} the tensors $B_{[n]}$ are related to $A_{[n]}(\theta_0)$ [thus becoming $\theta_0$-dependent, $B_{[n]}\to B_{[n]}(\theta_0)$] via the SVD procedure as described by Eq.~\eqref{eq:SVD_main}, allowing us to express 
\begin{equation}
\label{eq:Lambda_theta_0}
\Lambda(\theta_0,T) =\rho_{\rm S}(\theta_0, T)\left(R_{[N]}^\dag(\theta_0)\right)^{-1},
\end{equation}
in which 
\begin{align}
\rho_{\rm S} (\theta_0,T)=\sum _{\left\{\sigma _n\right\}} A_{[N]}^{\sigma _N}(\theta_0)\dots A_{[1]}^{\sigma _1}(\theta_0)|\psi _{\rm S}(0)\rangle \nonumber\\
\times\langle
\psi _{\rm S}(0)|A_{[1]}^{\sigma _1\dagger}(\theta_0)\dots A_{[N]}^{\sigma _N\dagger}(\theta_0)
\end{align}
is the sensor density matrix at time $T$.
We therefore have
\begin{equation}
P_{\theta_0}[\{\sigma_n^*\}]= {\rm tr}\left[\rho_{\rm S}^2(\theta_0,T)\left(R_{[N]}(\theta_0)R_{[N]}^\dag(\theta_0)\right)^{-1}\right]=1,
\label{eq:probability_dark}
\end{equation}
where we have used the fact that ${\rho}_{\rm S}(\theta_0,T)={R}_{[N]}(\theta_0)R_{[N]}^\dag(\theta_0)$ as shown in Sec.~\ref{sec:decoder_recipe}. 

\subsection{The Sensor-Decoder Entangled State}
\label{app:B2}
The above formal proof also allows us to analyze the sensor-decoder entanglement established by our information retrieval method, consolidating the presentation of Sec.~\ref{sec:sensor-decoder-entanglement}. To this end, we first notice, as per Eq.~\eqref{eq:R_general_solution}, that Eq.~\eqref{eq:Lambda_theta_0} can be rewritten as
\begin{equation}
\Lambda(\theta_0,T)=\sqrt{\rho_{\rm S}(\theta_0,T)} W(T),
\end{equation}
with $W(T)$ the time-dependent unitary reflecting the gauge redundancy. Adopting the spectrum decomposition $\rho_{\rm S}{(\theta_0,T)}=\sum_{k=1}^D p_k(\theta_0,T) \ket{k(\theta_0,T)}\bra{k(\theta_0,T)}$, we can express $\Lambda(\theta_0,T)$ as
\begin{equation}
\Lambda(\theta_0,T)=\sum_{k=1}^D \sqrt{p_k(\theta_0,T)} \ket{k(\theta_0,T)}\bra{k(\theta_0,T)}W(T).
\end{equation}
Using the isomorphism $|\tilde{\psi}_{\rm SD}(\theta,T)\rangle\leftrightarrow\Lambda(\theta,T)$, cf. Eqs.~\eqref{eq:psi_SD_evolve} and \eqref{eq:choi_map_matrix}, and noticing that $\ket{\psi_{\rm SD}(\theta_0,T)}$ is a normalized state, we conclude that the sensor and the decoder are in a pure entangled state for $\theta=\theta_0$, $\rho_{\rm SD}(\theta_0,T)=\ket{\psi_{\rm SD}(\theta_0,T)}\bra{\psi_{\rm SD}(\theta_0,T)}$, with
\begin{align}
\ket{\psi_{\rm SD}(\theta_0,T)}=&\sum_{k=1}^D \sqrt{p_k(\theta_0,T)} \ket{k(\theta_0,T)}_{\rm S} \nonumber\\
&\otimes\left(W^\dag(T)\ket{k(\theta_0,T)}_{\rm D}\right)
\end{align}
a natural purification of  the sensor density matrix $\rho_{\rm S}{(\theta_0,T)}$. For $\theta\neq\theta_0$ but sufficiently close to $\theta_0$, the sensor and decoder is in a mixed entangled state due to entanglement with the environment.

\subsection{The Effective Measurement is Optimal}
\label{app:optimal_proof}
We proceed to prove {\bf (I)}, i.e., the effective measurement that satisfies Eqs.~\eqref{eq:optimal_cond} and \eqref{eq:vac_projector} can fully retrieve the QFI of the emission field up to a finite constant. The retrieved FI, $F(\theta, T)$, is determined by the statistical distribution $P_{\theta}[D(0,T)]$ of the quantum trajectories of the stochastic cascaded ME~\eqref{eq:SME} for the joint sensor and decoder. For unknown parameter $\theta$ located in the neighborhood of the prior knowledge $\theta_0$, such distribution is provided via the Taylor expansion around $\theta_0$,
\begin{align}
\label{eq: boundary_condition_statistical_model}
P_{\theta}[\{\sigma_n^*\}]=&\,1-\frac{1}{2}(\theta-\theta_0)^2\partial_{\theta}^2P_{\theta}[\{\sigma_n^*\}]|_{\theta=\theta_0}\nonumber\\
&+{\cal O}[(\theta-\theta_0)^3],\nonumber\\
P_{\theta}[D(0,T)]=&\,\frac{1}{2}(\theta-\theta_0)^2\partial_{\theta}^2P_{\theta}[D(0,T)]|_{\theta=\theta_0}\nonumber\\
&+{\cal O}[(\theta-\theta_0)^3],\:\: \forall D(0,T)\neq\{\sigma_n^*\}, 
\end{align}
where $\{\sigma_n^*\}=\{0,0,\dots,0\}$ corresponds to the detection of null photons as defined in the main text, and `${\cal O}(x)$' stands for `on the order of $x$'. In arriving at Eq.~\eqref{eq: boundary_condition_statistical_model}, we have used $P_{\theta_0}[\{\sigma_n^*\}]=1$ as per Eq.~\eqref{eq:optimal_cond}, and the fact $\partial_\theta P_\theta[D(0,T)]|_{\theta=\theta_0}=0, \forall D(0,T)$, which is a necessary condition to ensure $0\leq P_\theta[D(0,T)]\leq 1$. As a result, 
\begin{align}
\label{eq:FI_theta0}
F(\theta,T)&=\sum_{D(0,T)}\frac{\{\partial_\theta P_{\theta}[D(0,T)]\}^2}{P_{\theta}[D(0,T)]}\nonumber\\
&=-2\partial_\theta^2 P_{\theta}[\{\sigma_n^*\}]|_{\theta=\theta_0}+{\cal O}(\theta-\theta_0),
\end{align}
where we have used the fact $\sum_{D(0,T)}P_{\theta}[D(0,T)]=1$. Equation \eqref{eq:FI_theta0} shows that the FI retrieved by our measurement scheme is continuous in the neighborhood of $\theta_0$, around the central value
\begin{align}
\label{eq:FI_theta0_new}
F(\theta_0,T)=&-2\partial_\theta^2 P_{\theta}[\{\sigma_n^*\}]|_{\theta=\theta_0}\nonumber\\
\equiv&-2\partial_\theta^2\Big(\langle\tilde{\psi}_{\rm SD}(\theta,T)|\tilde{\psi}_{\rm SD}(\theta,T)\rangle\Big)\Big|_{\theta=\theta_0},
\end{align}
where $\ket{\tilde{\psi}_{\rm SD}(\theta, T)}$ is defined in Eq.~\eqref{eq:psi_SD}. This establishes a simple link between the retrieved FI and the norm of the conditional state $\ket{\tilde{\psi}_{\rm SD}(\theta,T)}$. We emphasize that such a link is a direct consequence of Eqs.~\eqref{eq:optimal_cond} and \eqref{eq:vac_projector} that defines our effective measurement. 

Under Eqs.~\eqref{eq:optimal_cond} and \eqref{eq:vac_projector}, we can relate the two QFIs introduced in Sec.~\ref{sec:eQFI} with $\ket{\tilde{\psi}_{\rm SD}(\theta, T)}$ as well. It turns out that the relation with the global QFI, $I_{\rm G}(\theta,T)$, is particularly compact and provides sufficient physical insights. The global QFI is defined in Eq.~\eqref{eq:QFIg} in terms of the sensor-environment global state $\ket{\Psi(\theta,T)}$ [cf. Eq.~\eqref{eq:MPSglobal}] prior to the interaction with the decoder. Here we adopt an equivalent expression
\begin{align}
\label{eq:QFIg_alt}
I_{\rm G}(\theta,T)&=-2\partial_\delta^2 {\cal F}_{\rm G}^2(\theta,\theta+\delta)|_{\delta=0}\nonumber\\
&\equiv-2\partial_\delta^2\Big(|\langle\Psi(\theta,T)|\Psi(\theta+\delta,T)\rangle|^2\Big)\Big|_{\delta=0}.
\end{align}
The equivalence stems from the fact that if an analytic function $f(\delta)$ satisfies $f(\delta)\leq f(0)=1$, $\forall \delta\in \mathbb{R}$, then $\partial_{\delta}f|_{\delta=0}=0$ and $\partial_\delta^2 f^2|_{\delta=0}=2\partial_{\delta}^2f|_{\delta=0}$, as can be checked straightforwardly via the Taylor expansion of $f^2(\delta)$. By setting $f(\delta)={\cal F}_{\rm G}(\theta,\theta+\delta)$ we convert Eq.~\eqref{eq:QFIg} to Eq.~\eqref{eq:QFIg_alt}.

In our retrieval scheme, by coupling the decoder downstream from the sensor, the global state $\ket{\Psi(\theta,T)}$ is transformed into $\ket{\Psi_{\rm tot}(\theta, T)}=U_{\rm DE}(\theta_0,T) \ket{\Psi(\theta,T)}\otimes\ket{\psi_{\rm D}(0)}$ of the sensor, the environment and the decoder as defined in Appendix \ref{app: proof_left_normalization}. As $U_{\rm DE}(\theta_0,T)$ is independent of $\theta$, we are allowed to rewrite Eq.~\eqref{eq:QFIg_alt} as
\begin{equation}
I_{\rm G}(\theta, T)=-2\partial_\delta^2\Big(|\langle\Psi_{\rm tot}(\theta,T)|\Psi_{\rm tot}(\theta+\delta,T)\rangle|^2\Big)\Big|_{\delta=0}.
\label{eq:IG_tot}
\end{equation}
Let us focus on $\theta=\theta_0$, at which our effective measurement, as defined by Eqs.~\eqref{eq:optimal_cond} and \eqref{eq:vac_projector}, prescribes $\ket{\Psi_{\rm tot}(\theta_0,T)}=\ket{\rm vac}\otimes\ket{\psi_{\rm SD}(\theta_0,T)}$. Substituting this into Eq.~\eqref{eq:IG_tot} gives
\begin{align}
\label{eq:IGtheta0}
I_{\rm G}(\theta_0,T)&=-2\partial_\theta^2\Big(|\langle\tilde{\psi}_{\rm SD}(\theta,T)|\psi_{\rm SD}(\theta_0,T)\rangle|^2\Big)\Big|_{\theta=\theta_0}.
\end{align}
Equation \eqref{eq:IGtheta0} links the global QFI evaluated at $\theta=\theta_0$ to the conditional state $\ket{\tilde{\psi}_{\rm SD}(\theta,T)}$. This, again, is a direct consequence of Eqs.~\eqref{eq:optimal_cond} and \eqref{eq:vac_projector} that defines our effective measurement. 

Using Eqs.~\eqref{eq:FI_theta0_new} and \eqref{eq:IGtheta0}, we can express the difference between the retrieved FI and the global QFI as
\begin{equation}
\label{eq:IGF_dif}
I_{\rm G}(\theta_0,T)-F(\theta_0,T)
=\,2\,\langle\partial_\theta\tilde{\psi}_{\rm SD}(\theta,T)|\mathbb{P}|\partial_\theta\tilde{\psi}_{\rm SD}(\theta,T)\rangle|_{\theta=\theta_0}
\end{equation}
with $\mathbb{P}=1-\ket{\psi_{\rm SD}(\theta_0,T)}\bra{\psi_{\rm SD}(\theta_0,T)}$ a projector. We note that the right-hand-side of Eq.~\eqref{eq:IGF_dif} recovers (up to a factor of 2) the familiar pure-state QFI formula $4[\langle\partial_\theta\phi|\partial_\theta\phi\rangle-|\langle\phi|\partial_\theta\phi\rangle|^2]$, if we replace the unnormalized conditional state $\ket{\tilde{\psi}_{\rm SD}(\theta,T)}$ with a nomalized one $\ket{\phi(\theta,T)}$. As such, we can loosely interpret this term as the `leftover' information stored in the sensor and the decoder that our effective measurement can not retrieve. The conditional state $\ket{\tilde{\psi}_{\rm SD}(\theta,T)}$ evolves according to the sequential map Eq.~\eqref{eq:psi_SD_evolve} which, in the continuous time limit, is equivalent to the SME~\eqref{eq:SME} conditioned upon the detection of null photons, $d{\cal N}(t)=0, \forall t\in[0,T]$. Such an evolution is in general not trace preserving---the norm of $\ket{\tilde{\psi}_{\rm SD}(\theta,T)}$ decays (typically exponentially) with time for $\theta\neq\theta_0$. Only at $\theta=\theta_0$, the evolution is trace preserving---we have shown in Appendix~\ref{app: proof_left_normalization} that it can be related to the Kraus maps describing the sensor evolution. As we show below, it is exactly such dissipative, non-unitary evolution that dictates Eq.~\eqref{eq:IGF_dif} typically becomes a constant for long evolution time $T$. In stark contrast, if we replace $\ket{\tilde{\psi}_{\rm SD}(\theta,T)}$ in Eq.~\eqref{eq:IGF_dif} with a nomalized state $\ket{\phi(\theta,T)}$ that evolves unitarily, it is well known that the corresponding pure-state QFI can (and typically does) grow with time. 

To this end, let us define the $n$th map in Eq.~\eqref{eq:psi_SD_evolve} as ${\cal K}_{[n]}(\theta):=\sum_{\sigma_n}\bar{B}_{[n]}^{\sigma_n}(\theta_0)\otimes A_{[n]}^{\sigma_n}(\theta)$, and ${\cal K}(\theta,m,l):={\cal T}\prod_{n=m+1}^{l}{\cal K}_{[n]}(\theta)$. This allows us to re-express Eq.~\eqref{eq:psi_SD_evolve} as 
\begin{equation}
\label{eq:psi_SD_alt}
\ket{\tilde{\psi}_{\rm SD}(\theta,T)}={\cal K}(\theta,0,N)\ket{\psi_{\rm D}(0)}\otimes\ket{\psi_{\rm S}(0)}
\end{equation}
and as a result 
\begin{align}
\label{eq:partial_psiSD_general}
\partial_{\theta}\ket{\tilde{\psi}_{\rm SD}(\theta,T)}|_{\theta=\theta_0}=&\sum_{m=1}^{N}{\cal K}(\theta_0,m,N)\nonumber\\
&\left(\partial_\theta {\cal K}_{[m]}(\theta)|_{\theta=\theta_0}\right)\ket{\psi_{\rm SD}(\theta_0,t_{m-1})},
\end{align}
where $t_m\equiv m dt$ as defined in Sec.~\ref{sec:coarse-graining}, and we have made the convention that ${\cal K}(\theta_0,N,N)=\mathbb{I}_{D^2}$. As discussed in Appendix~\ref{app: proof_left_normalization}, our effective measurement condition Eqs.~\eqref{eq:optimal_cond} and \eqref{eq:vac_projector} prescribes that the map ${\cal K}(\theta_0,m,N)$ satisfy
\begin{equation}
\label{eq:K_SD}
{\cal K}(\theta_0,m,N)=(\bar{R}_{[N]}^{-1}\otimes \mathbb{I}_{D}){\cal M}(\theta_0,m,N)(\bar{R}_{[m]}\otimes \mathbb{I}_{D})
\end{equation}
with
\begin{align}
\label{eq:Lsensor}
{\cal M}(\theta_0,m,N)=&\sum_{\{\sigma_n\}}\left(\bar{A}_{[N]}^{\sigma_{N}}(\theta_0)\dots\bar{A}_{[m+1]}^{\sigma_{m+1}}(\theta_0)\right)\nonumber\\
&\otimes \left(A_{[N]}^{\sigma_N}(\theta_0)\dots A_{[m+1]}^{\sigma_{m+1}}(\theta_0)\right)
\end{align}
being the Choi-Jamiolkowski isomorphism of the Kraus map describing the sensor evolution in the time interval $[t_m,T]$. Being a CPTP map, ${\cal M}(\theta_0,m,N)$ has at least one fixed point (i.e., stationary state)~\cite{breuer2002theory,Rivas_2012}. As we show below, the large-$T$ behavior of the `leftover' information Eq.~\eqref{eq:IGF_dif} is strongly dependent on whether such a fixed point is unique---that is, whether the sensor evolution is \emph{ergodic}.

To keep our presentation easy to follow, we will first illustrate this aspect with sensors whose evolution is time homogeneous [that is, the tensors in Eq.~\eqref{eq:Lsensor} are identical, $A_{[n]}\equiv A$], for which ergodicity is a well-established mathematical property, see, e.g., Ref.~\cite{bremaud2001markov}. Note that most driven-dissipative sensors in the laboratory are time-stationary (and are moreover ergodic). We will then extend our discussion to sensors whose evolution is time-inhomogeneous, invoking the mathematical concept of `strong ergodicity'~\cite{bremaud2001markov}. Note that for the time-inhomogeneous case (corresponding to the so-called inhomogeneous Markov chain in mathematics), a complete understanding of ergodicity and related properties is yet to be established~\cite{bremaud2001markov}.

\subsubsection{Time-homogeneous sensor evolution}
\label{sec:homo_proof}
In this case $A_{[n]}(\theta_0)=A(\theta_0)$, and the fixed point of the map ${\cal M}(\theta_0,m,N)$ is identical to the fixed point of the elementary map $\sum_{\sigma_n}\bar{A}^{\sigma_n}(\theta_0)\otimes A^{\sigma_n}(\theta_0):={\cal M}(\theta_0)$ of each time-bin. If the fixed point is unique, the map ${\cal M}(\theta_0)$ [and thus ${\cal M}(\theta_0,m,N)$] is \emph{ergodic}; correspondingly, the MPS constructed from $A(\theta_0)$ [cf. Eq.~\eqref{eq:MPSglobal}] is an \emph{injective} representation~\cite{RevModPhys.93.045003}. Otherwise, ${\cal M}(\theta_0)$ is \emph{nonergodic}; the corresponding MPS is not injective. 

Let us first focus on the ergodic case by assuming that ${\cal M}(\theta_0)$ has a unique fixed point $\ket{\rho_{\rm S}^{\rm st}(\theta_0)}$, ${\cal M}(\theta_0)\ket{\rho_{\rm S}^{\rm st}(\theta_0)}=\ket{\rho_{\rm S}^{\rm st}(\theta_0)}$, which is the isomorphic to the stationary density matrix $\rho^{\rm st}_{\rm S}(\theta_0)$ of the sensor. We denote (the real part of) the second largest eigenvalue of ${\cal M}(\theta_0)$ as $e^{-\xi}$ with $\xi>0$. This allows us to write down the spectrum decomposition
\begin{equation}
\label{eq:Ltheta}
{\cal M}(\theta_0,m,N)=\ket{\rho_{\rm S}^{\rm st}(\theta_0)}\bra{\mathbb{I}_D} + {\cal O}[e^{-\xi(N-m)}],
\end{equation}
where $\ket{\mathbb{I}_{D}}$ is the Choi isomorphism of $\mathbb{I}_D$. Correspondingly, Eq.~\eqref{eq:K_SD} can be converted to
\begin{align}
\label{eq:Ltheta_alt}
{\cal K}(\theta_0,m,N)=&(\bar{R}_{[N]}^{-1}\otimes \mathbb{I}_{D})\ket{\rho_{\rm S}^{\rm st}(\theta_0)}\bra{\mathbb{I}_D}(\bar{R}_{[m]}\otimes \mathbb{I}_{D}) \nonumber\\
&+ {\cal O}[e^{-\xi(N-m)}].
\end{align}
This indicates, via Eq.~\eqref{eq:psi_SD_alt}, that $\ket{{\psi}_{\rm SD}(\theta_0,T)}=(\bar{R}_{[N]}^{-1}\otimes \mathbb{I}_{D})\ket{\rho_{\rm S}^{\rm st}(\theta_0)}+{\cal O}[e^{-\xi(N-m)}]$, and therefore
\begin{equation}
\label{eq:K_alt_homo}
{\cal K}(\theta_0,m,N)=\ket{{\psi}_{\rm SD}(\theta_0,T)}\bra{\mathbb{I}_D}(\bar{R}_{[m]}\otimes \mathbb{I}_{D}) +{\cal O}[e^{-\xi(N-m)}].
\end{equation}
Physically, this means that if the sensor evolution is ergodic, so is the sensor-decoder evolution constructed by our retrieval scheme. Inserting Eq.~\eqref{eq:K_alt_homo} into Eqs.~\eqref{eq:partial_psiSD_general} and \eqref{eq:IGF_dif}, we arrive at
\begin{align}
\label{eq: laeq}
I_{\rm G}(\theta_0,T)-F(\theta_0,T)=&\left(\sum_{m=1}^N{\cal O}\left[e^{-\xi(N-m)}\right]\right)\nonumber\\
&\times\left(\sum_{n=1}^N {\cal O}\left[e^{-\xi(N-n)}\right]\right)\nonumber\\
=&\,{\rm const.}
\end{align}
The second equality of Eq.~\eqref{eq: laeq} can be verified easily by taking the continuous-time limit, where each bracket in the second line becomes $\sim \int_0^T dt {\rm exp}[-\xi (T-t)/\Delta t]\sim {\rm const}$. 
Thus, we have rigorously proven that our retrieval scheme is optimal for all sensors whose evolution is time-homogeneous and ergodic. 

Next, let us look at the case that the sensor evolution has multiple stationary states, $\rho^{\rm st}_{\rm S}(\theta_0,\alpha)$, $\alpha=1,2,...d$. Correspondingly, the sensor density matrix at time $T$ can be expanded as $\rho_{\rm S}(\theta_0,T)=\sum_{\alpha}\upsilon_\alpha \rho^{\rm st}_{\rm S}(\theta_0,\alpha)+{\cal O}(e^{-\xi N})$. Repeating the calculation above, we find
\begin{align}
\label{eq:Ltheta_multi}
I_{\rm G}(\theta_0,T)-F(\theta_0,T)=\sum_{m,n=1}^N\sum_{\alpha,\beta} c_{n,\alpha}^*c_{m,\beta}\left(\frac{1}{\upsilon_\beta}-\upsilon_\alpha\upsilon_\beta\right)&\nonumber\\
+\sum_{m,n}^N{\cal O}\left[e^{-\xi(N-m)}\right]{\cal O}\left[e^{-\xi(N-n)}\right]&
\end{align}
where we have defined the coefficients $c_{n,\alpha}:=\langle \mathbb{I}_{D}|\left(\bar{R}_{[m]}\otimes \mathbb{I}_{D}\right)(\partial_\theta {\cal K}_{[m]}(\theta)|_{\theta=\theta_0})\ket{\psi_{\rm SD}(\theta_0,t_{m-1})}$. While the second line of Eq.~\eqref{eq:Ltheta_multi} contribute to a constant, the first term is, in general, unbounded in time. Physically, this is due to the fact that the emission field MPS generated by non-ergodic open dynamics may not be injective and may possess long-range correlation. Indeed, for such a case not only $I_{\rm G}(\theta_0,T)-F(\theta_0,T)$, but also $I_{\rm E}(\theta_0,T)-F(\theta_0,T)$ may grow with time. For the three-level sensor example presented in Sec.~\ref{sec:3LS}, in the time window $[0,T]$ where $\Omega_{\rm S,1}$ is a constant, the sensor evolution is time-homogeneous and has two stationary states. Correspondingly, $I_{\rm G}(\theta_0,T)-F(\theta_0,T)$ increases with time [for this specific model $I_{\rm E}(\theta_0,T)-F(\theta_0,T)$ is a finite constant as verified by our numerics; but it may grow with time for general non-ergodic models]. The application of the short pulse $\Omega_{\rm S,2}(t)$ and subsequent relaxation of the sensor modifies the sensor dynamics, leading to a unique stationary state at the end of the interrogation. This allows for the efficient retrieval of the Heisenberg-limited (global) QFI.

\subsubsection{Time-inhomogeneous sensor evolution}
Let us now extend the above analysis to time-inhomogeneous driven-dissipative dynamics. As the tensors $A_{[n]}$ are inhomogeneous, the fixed point of the map ${\cal M}(\theta_0,m,N)$, cf. Eq.~\eqref{eq:Lsensor}, is no longer determined by the fixed point of the evolution of each time-bin. Indeed, in the most general case the fixed point of ${\cal M}(\theta_0,m,N)$ may differ for different $m$. A proper notion of ergodicity in inhomogeneous classical Markov chains invokes `strong ergodicity', see, e.g., Chapter 12 of Ref.~\cite{bremaud2001markov}; direct translation to the quantum case leads to

\emph{\textbf{Definition} (Quantum Strong Ergodicity).} The set of CPTP maps ${\cal M}(\theta_0,m,N)$ in Eq.~\eqref{eq:Lsensor} is \emph{strongly ergodic} iff there exists a unique $\ket{\rho_{\rm S}^{\rm st}(\theta_0)}$ such that
\begin{equation}
{\rm lim}_{N\to \infty} {\cal M}(\theta_0,m,N)\ket{\rho_{\rm S}}=\ket{\rho_{\rm S}^{\rm st}(\theta_0)}
\end{equation}
for all $m$ and all (normalized) initial state $\ket{\rho_{\rm S}}$.

Invoking such a definition, we will see below that the proof in Appendix \ref{sec:homo_proof} can be largely extended to the time-inhomogeneous case, thus validating the optimum of our retrieval scheme also for time-dependent sensor designs. To this end, we again write down the spectrum decomposition of ${\cal M}(\theta_0,m,N)$, akin to Eq.~\eqref{eq:Ltheta}, under the assumption of strong ergodicity, 
\begin{equation}
\label{eq:Ltheta2}
{\cal M}(\theta_0,m,N)=\ket{\rho_{\rm S}^{\rm st}(\theta_0)}\bra{\mathbb{I}_D} + {\cal O}({\chi_m}),
\end{equation}
where $\chi_m$ is the the real part of the second largest eigenvalue of ${\cal M}(\theta_0,m,N)$. $\chi_m$ generally decays with the evolution time $N-m$ to ensure ergodicity, but not necessarily exponentially as in Eq.~\eqref{eq:Ltheta}. On the basis of Eq.~\eqref{eq:Ltheta2}, it is straightforward to verify that Eq.~\eqref{eq:Ltheta_alt} and Eq.~\eqref{eq:K_alt_homo}, with the replacement ${\cal O}[e^{-\xi(N-m)}]\to {\cal O}({\chi_m})$, also hold for the inhomogeneous case. As a result, we arrive at
\begin{equation}
I_{\rm G}(\theta_0,T)-F(\theta_0,T)=\sum_{m,n=1}^N{\cal O}(\chi_m){\cal O}(\chi_n),
\end{equation}
which, apparently, is a constant as long as $\chi_m$ decays with the evolution time fast than $\sim (N-m)^{-1}$. This is a very mild requirement fulfilled by most sensor designs under smooth time-dependent driving protocols. We thus have extended the effectiveness of our retrieval scheme to sensors whose evolution is ergodic despite being time-inhomogeneous.

In contrast, general time-inhomogeneous, non-ergodic evolution may result in long-range correlation in the emission field, for which the efficiency of our retrieval scheme is not guaranteed---not only $I_{\rm G}(\theta_0,T)-F(\theta_0,T)$, but also $I_{\rm E}(\theta_0,T)-F(\theta_0,T)$ may grow with time in this case.

\subsection{Extension to Discrete Matrix Product States}
\label{app:discreteMPS}
The proof presented above, thus the effectiveness of our information retrieval scheme, holds equally for continuous (as relevant to the emission field of open quantum sensors) and discrete (as relevant to many-body lattice models) MPSs. As an illustration of the latter, let us consider a one-dimensional lattice spin model defined on $n=1,\dots, N$ sites with open boundary condition (as relevant to realistic experimental setups). We assume that the ground state of the spin chain is described by an MPS 
\begin{equation}
\ket{\Psi}=\sum_{\{\sigma_n\}}V_{[1]}^{\sigma_1}\dots V_{[n]}^{\sigma_n}\dots V_{[N]}^{\sigma_N}\ket{\sigma_1\dots\sigma_n\dots\sigma_N}.\nonumber
\end{equation}
Note that we label the sites from left to right---in contrast to the main text but in accord with the standard convention for discrete MPS~\cite{RevModPhys.93.045003}.
Without loss of generality, we assume that the tensors $V_{[n]}$ have a fixed bond dimension $D$; in the bulk, they become translational invariant, $V_{[n]}=V$, $1\ll n\ll N$, and we assume injectivity~\cite{RevModPhys.93.045003}. We consider that the unknown parameter $\theta$ (the corresponding prior information is $\theta_0$) enters the tensor $V_{[n]}(\theta)$ as local field, and we have access to a subsystem consisting of $n=1,2,...,L$ spins with $L\gg 1$.

To efficiently retrieve the QFI of the subsystem, we can apply our information retrieval scheme. We first convert the tensors $V_{[n]}(\theta_0)$, $n=1,2,...,L$  to the left normalized form $A_{[n]}(\theta_0)$ [which satisfy $\sum_{\sigma_n}A_{[n]}^{\sigma_n\dag}(\theta_0)A_{[n]}^{\sigma_n}(\theta_0)={\mathbb I}_D$] via sequential SVDs on a classical computer. We then use a $D$-dimensional ancilla to sequentially interact with each spin $1,\dots,L$. The unitary operator between the ancilla and the $n$-th spin $U_{[n]}(\theta_0)$ is chosen to satisfy 
\begin{align}
\label{eq:discMPS_choice}
A_{[n]}^{0}(\theta_0)=\bra{\downarrow_n}U_{[n]}^\dag(\theta_0)\ket{\downarrow_n},\nonumber\\
{A}_{[n]}^{1}(\theta_0)=\bra{\uparrow_n}U_{[n]}^\dag(\theta_0)\ket{\downarrow_n},
\end{align}
for $n=2,3,...,L$. For $n=1$, $U_{[n]}^\dag$ in Eq.~\eqref{eq:discMPS_choice} should be replaced by $\langle \psi_{\rm A}(0)|U_{[1]}^\dag$ with $\ket{\psi_{\rm A}(0)}$ the initial state of the ancilla. 
Finally, we perform site-resolved projective measurement of the subsystem of spins in the basis $\{\ket{\uparrow}_n,\ket{\downarrow}_n\}$. Our studies in this work guarantee that such a measurement is optimal---it can retrieve the QFI of the subsystem up to a finite constant which, importantly, does not grow with the subsystem size $L$.

The sequential interaction of an ancilla with multiple spins has been demonstrated in trapped ion quantum simulators~\cite{PhysRevX.12.041035}, and may also be possible with circuit QED setups and nitrogen-vacancy centers in diamond.

\subsection{Efficient Processing of the Measurement Data}
\label{app:B5}
When our information retrieval scheme is applied to general open sensors, the statistics of the resultant measurement data is described by the distribution Eq.~\eqref{eq: boundary_condition_statistical_model} in the neighborhood of the prior information $\theta_0$. Such a statistical model also appears in various scenarios of metrology, e.g., in super-resolution astronomy~\cite{PhysRevX.6.031033}. For such a statistical model, it has been shown that the amplitude of the small deviation, $|\theta-\theta_0|$, can be efficiently determined by maximum likelihood estimation (MLE)~\cite{PhysRevX.6.031033}. In general, the sign of $\theta-\theta_0$ can not be efficiently identified by the statistical model alone, as Eq.~\eqref{eq: boundary_condition_statistical_model} depends only on $(\theta-\theta_0)^2$ to the lowest order Taylor expansion. Nevertheless, the sign can be predetermined using the prior information~\cite{madalin}: Suppose the prior information provides a confidence interval $\theta\in(\vartheta_0-\epsilon,\and\vartheta_0+\epsilon)$, one can deliberately choose the decoder parameter $\theta_0$ to be outside this interval, e.g., $\theta_0=\vartheta_0\,+$ a few $\epsilon$, thus fixing the sign of $\theta-\theta_0$. One can then determine $\theta-\theta_0$ via counting the decoder output, of which the (normalized) inverse variance converges to $F(\theta,T)\simeq F(\theta_0,T)$, provided $\epsilon$ is sufficiently small. A rigorous formulation of these aspects will be reported in a forthcoming article by M. Guta and coauthors~\cite{madalin}.

It is interesting to point out that if $\theta$ is located exactly at $\theta_0$, the FI is not a good quantifier of the estimation error~\cite{PhysRevX.6.031033,cox1979theoretical}. Indeed, in this case the estimator of MLE always coincides with the actual value (and therefore the associated estimation error is zero), a statistical phenomenon called `super-efficiency'~\cite{PhysRevX.6.031033, cox1979theoretical}.

\section{Calculation of the Decoder Parameters for Linear Sensors}
\label{app: linear_sensor}
Without loss a generality, we assume that both the sensor and the decoder are initialized in the ground state of the oscillator. The driven dissipative evolution of a linear sensor preserves the Gaussianity of its density matrix. The sensor state $\rho_{\rm S}(t)$ can therefore be parameterized as $\rho_{\rm S}(t)={\rm exp}\{-[Q-u(t)]^{\rm T}G(t)[Q-u(t)]\}$, where $Q=(X_{\rm S}, P_{\rm S})$ consists of the two orthogonal quadrature operators of the oscillator, $u(t)=(\langle X_{\rm S}(t)\rangle,\langle P_{\rm S}(t)\rangle)$ with $\langle O(t)\rangle:={\rm tr}[O\rho_{\rm S}(t)]$, and $G(t)$ is related to the covariance matrix $C(t)$ of the Gaussian state via $G=2\sigma_y {\rm cot}^{-1}(2 C(t)\sigma_y)$~\cite{PhysRevLett.115.260501}, with $\sigma_y$ the $y$-component of the Pauli operators. Such a representation allows us to construct $\sqrt{\rho_{\rm S}(t)}$ conveniently via the replacement $G(t)\to G(t)/2$ in $\rho_{\rm S}(t)$. Straightforward calculation provides the mean quadratures
\begin{equation}
u(t)=\left(\frac{f}{m\omega^2}(1-\cos \omega t),\frac{f}{\omega}\sin\omega t\right)\nonumber
\end{equation}
and the covariance matrix elements
\begin{align}
C_{XX}(t)&=\frac{1}{2m\omega}+\frac{\Gamma}{2m^2\omega^2}\left(t-\frac{1}{2\omega}\sin\omega t\right),\nonumber\\
C_{PP}(t)&=\frac{1}{2}m\omega+\frac{\Gamma}{2}\left(t+\frac{1}{2\omega}\sin\omega t\right),\nonumber\\
C_{XP}(t)&=C_{PX}(t)=\frac{\Gamma}{4m\omega^2}\left(1-\cos2\omega t\right).\nonumber
\end{align}
The use of these expressions and Eq.~\eqref{eq:H_J_general_solution} provide us with the decoder parameters as captured by Eq.~\eqref{eq:negmassH}.

\bibliography{references}

\end{document}